\documentclass[10pt]{article}
\usepackage{lscape}

\usepackage{amsfonts}
\usepackage{amsmath}
\usepackage{amsthm}
\usepackage{amssymb}
\usepackage{graphicx}
\usepackage{textcomp}

\def\be{\begin{eqnarray}}
\def\ee{\end{eqnarray}}
\def\nn{\nonumber}

\def\a{\alpha}

\def\Tr{{\rm Tr}\,}

\def\l[{\phantom.[}

\hoffset= - 1.0in         

\begin{document}

\title{{\bf {Colored knot polynomials.
\\ HOMFLY in representation $[2,1]$
}\vspace{.2cm}}
\author{{\bf A. Mironov$^{a,b,c,d,}$}\footnote{mironov@lpi.ru; mironov@itep.ru}, \ {\bf A. Morozov$^{b,c,d,}$}\thanks{morozov@itep.ru},
 \ {\bf An. Morozov$^{b,c,d,e,}$}\thanks{Andrey.Morozov@itep.ru} \ and {\bf
A. Sleptsov$^{b,c,d,e,}$}\footnote{sleptsov@itep.ru}}
\date{ }
}

\maketitle

\vspace{-6.0cm}

\begin{center}
\hfill FIAN/TD-5/15\\
\hfill IITP/TH-10/15\\
\hfill ITEP/TH-17/15\\
\end{center}

\vspace{4.2cm}

\begin{center}
$^a$ {\small {\it Lebedev Physics Institute, Moscow 119991, Russia}}\\
$^b$ {\small {\it ITEP, Moscow 117218, Russia}}\\
$^c$ {\small {\it Institute for Information Transmission Problems, Moscow 127994, Russia}}\\
$^d$ {\small {\it National Research Nuclear University MEPhI, Moscow 115409, Russia }}\\
$^e$ {\small {\it Laboratory of Quantum Topology,
Chelyabinsk State University, Chelyabinsk 454001, Russia}}
\end{center}

\vspace{1cm}

\begin{abstract}
This paper starts a systematic description of colored
knot polynomials, beginning from the first non-(anti)symmetric representation $R=[2,1]$.
The project involves several steps:

(i) parametrization of big families of knots {\it a la} \cite{MM3str},

(ii) evaluating Racah/mixing matrices for various numbers of strands
in various representations {\it a la} \cite{MMMknots2},

(iii) tabulating and collecting the results at \cite{ruskatlas}.

\noindent
In this paper we discuss only representation $R=[2,1]$ and construct all necessary ingredients that allow one to
evaluate knot/links represented by three strand closed parallel braids with inserted double-fat fingers. In particular,
it is used to evaluate knots from a 7-parametric family: this family contains over
80\% of knots with up to 10 intersections, but does not include mutants.
\end{abstract}

\vspace{.5cm}

\section{Introduction}

Knot polynomials \cite{knotpol} are among the hottest subjects of modern theory, interesting both for
physics and mathematics.
This is because they are correlators of Wilson lines in $3d$
Chern-Simons (CS) theory \cite{CS,Wit}  (sometimes deformed)
and are naturally related to $2d$ conformal blocks, both with extended symmetry (WZWN models) and without it, \cite{Wit,inds,GMM,GalMelMM,Fabio}, to matrix models and tau-functions \cite{MAMOtau,tau,OV,MMMknots1,MMS2} -- the main types of
special functions we currently know and use.
Therefore, it is not a surprise that knot polynomials appear in quantitative description
of various seemingly unrelated problems, from augmentation varieties in Calabi-Yau manifolds \cite{Aug}
to the theory of quantum Hall effect.
They satisfy a vast variety of linear and non-linear relations, of which we yet discovered
only a small fraction \cite{Gar,MMeqs}, and this makes us to believe that complete theory of knot polynomials
is actually available.

The most important HOMFLY polynomials
\be
H^{{\cal L}\subset {\cal M}}_R(q,A) = \ \left< {\rm Tr}_R \ P\exp\oint_{\cal L} {\cal A} \right>^{CS}
\ee
depend on five types of arguments: the knot/link ${\cal L}$ in a three-dimensional space ${\cal M}$,
the CS coupling constant $q = \exp\left(\frac{2\pi i}{k+N}\right)$,
the Lie algebra $G=SU(N)$ with $A=q^N$
and its representation (Young diagram) $R$.
Similar polynomials  can be defined for other Lie algebras and groups,
they are usually related to HOMFLY polynomials \cite{KaufviaHOMFLY}.
These quantities are called polynomials, because, being properly normalized they are Laurent polynomials in variables $q$ and $A$,
at least when
${\cal M}$ is simply-connected: ${\cal M}=R^3 \ \text{or} \ S^3$.
Polynomiality is preserved for virtual knots \cite{virt,virtHOMFLY}.

Today, of most interest is taming the dependencies of HOMFLY polynomials on ${\cal L}$ and $R$.
For this purpose, a number of methods was developed to describe their dependencies on $A$ and $q$
for given ${\cal L}$ and $R$, and the goal of this and the subsequent papers in this series
is to find and list these dependencies at \cite{ruskatlas}.

\subsection{Modern versions of RT formalism}

The most effective approach to evaluate HOMFLY polynomials is based on various versions
of the Reshetikhin-Turaev (RT) approach \cite{RTfirst},\cite{inds},\cite{MMMknots2},\cite{IMMMfe}-\cite{mmmrs},\cite{MM3str}.
Its starting point is substitution of the link by its link diagram, a (2,2)-valent oriented
graph on a plane with two types of vertices (black and white), which is planar for ordinary
links and knots (but is non-planar for virtual ones).
After that, one substitutes the black and white vertices with the quantum ${\cal R}$-matrix of the quantum group $SU_q(N)$ and the inverse ${\cal R}$-matrix respectively
and contracts the indices, with additional weights $q^{\rho}$ per each Seifert cycle
(this procedure requires a seemingly small, but not fully understood modification for
virtual knots, where currently the only applicable is a very different hypercube method of \cite{Kh,DM3,virtHOMFLY}).
Despite the universal ${\cal R}$-matrices are long available, their convolutions and traces in the concrete representation are
not so easy to evaluate and one needs additional tricks to do this in a computationally effective way.

The key fact is that in each irreducible representation $Q$ in the product
\be
R_1\otimes R_2 = \ \oplus_Q\  W_Q\otimes Q
\label{decoR1R2}
\ee
the quantum ${\cal R}$-matrix acts proportionally to the identity operator,
with eigenvalues $\epsilon_Q q^{\varkappa_Q}$ where
$\varkappa_Q = \sum_{(i,j)\in Q} (i-j)$ and the sign factor $\epsilon_Q$ is plus or minus. For $R_1=R_2$ it is plus or minus
depending on wether $Q$ belongs to symmetric and antisymmetric square of $R$
respectively. Note that because of these sign factors $Q$ can enter (\ref{decoR1R2}) twice,
as $Q_+$ and $Q_-$, and the intertwiner spaces $W_{Q_+}$and $W_{Q_-}$ are {\it not} unified.

As a corollary of this general feature
in the particular case of the fundamental representation, $R_1=R_2=[1]=\Box$,
the ${\cal R}$-matrix satisfies the quadratic equation
\be
\Big({\cal R}-q\Big)\Big(q{\cal R}+1\Big) = 0 \ \ \
\Longleftrightarrow \ \ \
{\cal R} - {\cal R}^{-1} = q-q^{-1}
\ee
Changing normalization of the ${\cal R}$-matrix\footnote{In fact, this changing of normalization leads to the HOMFLY polynomials
in the topological framing, while the normalization that comes from the universal ${\cal R}$-matrix and is consistent with the group theory structure corresponds to the vertical framing.
}, one arrives at the skein relation \cite{skein},
\be\label{skein}
\Big(A{\cal R}_t-q\Big)\Big(Aq{\cal R}_t+1\Big) = 0 \ \ \
\Longleftrightarrow \ \ \
A{\cal R}_t - A^{-1}{\cal R}_t^{-1} = q-q^{-1}
\ee
which is very effective for evaluating the HOMFLY polynomials in the fundamental representation,
however a more sophisticated {\it cabling} procedure is required in other representations,
which becomes effective \cite{AnoMcabling} only in combination with other insights.

As soon as the ${\cal R}$-matrix acts trivially on the whole irreducible representation, one can consider it as acting
on the space of intertwining operators. From now on, we keep the same notation ${\cal R}$ for the ${\cal R}$-matrices acting on these spaces. These ${\cal R}$-matrices can act non-trivially only when these spaces are not one-dimensional.

The next simplification occurs when one considers fragments of link diagrams, which are braids.
Acting on the $m$-strand braid are $m-1$ different ${\cal R}$-matrices: ${\cal R}_{i,i+1}$
stands at the intersection of the strands $i$ and $i+1$.
If one diagonalizes ${\cal R}_{1,2}$, then ${\cal R}_{i,i+1} = {\cal U}_i{\cal R}_{1,2}{\cal U}_i^{-1}$
and ${\cal U}_i$ are called mixing matrices \cite{MMMknots2}. Clearly,
${\cal U}_1 = Id$, and ${\cal U}_2$ are just the Racah matrices, while higher ${\cal U}_i$
are complicated convolutions of various Racah matrices (see \cite{GalMelMM} for some examples).
The only case when an explicit description of all ${\cal R}_{i,i+1}$ is currently available,
is the fundamental representation, when they are represented by beautiful sums over paths in
the representation tree \cite{Anopaths},
which provides a nice a competitive alternative to the skein-relation technique.
Moreover, cabling is far more effective, if based on this formalism \cite{AnoMcabling},
still it remains a somewhat artificial procedure to describe non-fundamental representations.
It would be very desirable to find a more straightforward representation theory approach
to the problem.

As an example,
in \cite{eigen} (see also \cite{mathmix}), it was conjectured that in general
the entries of ${\cal U}_i$
are actually made from the eigenvalues of the corresponding ${\cal R}$. We partly use
this conjecture in sec.\ref{eigenvhyp} below.

\subsection{The present paper}

The main task of the present paper is to start a systematic investigation of the Racah
matrices in different representations.
We do this by a brute force method of \cite{MMMknots2}, just evaluating the highest weights
of representations $Q$, made from those of representations $R_i$ in two ways:
\be
(R_1\otimes R_2) \otimes R_3 =\ \oplus_{Q }\ {\cal W}_{Q } \otimes Q
\ee
and
\be
R_1\otimes (R_2 \otimes R_3) =\ \oplus_Q\ {\cal W}_Q' \otimes Q
\ee
and rotating one set of the highest weights for the given $Q$ into another.
In this particular paper we concentrate on the case of $R_1=R_2=R_3=[2,1]$,
the simplest one where the Racah matrices are unknown.

In a recent breakthrough paper \cite{GJ} slightly different matrices were found,
when either $R_2$ or $R_3$ are conjugate $\overline{[2,1]}$
(they are named respectively $\bar S$ and $S$ in \cite{mmmrs} and \cite{MM3str}),
so that the matrices essentially depend on $N$, -- but instead only the contributions
with $Q=[2,1]$ were picked up at the r.h.s., which is a great simplification as compared
to what we do in the present text.

While the results of \cite{GJ} were sufficient to consider highly non-trivial
examples of $[2,1]$-colored HOMFLY in \cite{nrs} and \cite{mmmrs}, including some mutants,
mixing this with the knowledge of arbitrary Racah matrices in the inclusive channel \ \
$[2,1]^3 \longrightarrow {all}$ \ \ allows one to do the same for entire families from
\cite{MM3str}, which are targeted at (and almost capable to) exhaustively describing
all knots with restricted number of intersections.
Of course, there is no way to present such a wide set of data in a paper,
these concrete results will be gradually posted at \cite{ruskatlas},
which is supposed to complement \cite{katlas} and \cite{virtkatlas}
by lists of colored polynomials and
their differential expansions in the style of \cite{evo,arthdiff}.

\subsection{Knot/link families}

There is still no systematic classification of knots and links,
different from a somewhat arbitrary enumeration in the Rolfsen tables
\cite{katlas,virtkatlas}.
In \cite{ruskatlas}, we also begin to use classification by families,
suggested in \cite{evo,mms,mmmrs} and \cite{MM3str}.

The old idea is that every knot and link has a closed braid representation,
thus one can study knot polynomials as functions of the number of strands
and brain words, what can be effectively done by the evolution method \cite{DMMSS,evo,GMMMS,mms}.
Technical problem here is that the minimal needed number of strands
can be pretty big for rather simple knots (like twist knots with many twists).
This is also a sign of the bad correlation between the minimal intersection
number in the link diagram and the minimal number of strands in the
braid representation.

Another idea \cite{prev,NRZ2,mmmrs} is to represent the knots and links
as the two-bridge "fingers" and "propagators" \cite{mmmrs}. These two ideas were recently combined \cite{MM3str}:
the two-bridge "fingers" and "propagators" can be attached to closed braids with a low number of strands. Already two \cite{mmmrs}
and three \cite{MM3str} strands provide very big families.
This looks like a far-going generalization of the pretzel family, studied in
\cite{GMMMS,mms,GalMelMM}, and is equally effective: it should be possible
to provide the answers for generic (anti)symmetric representations, for $R=[2,1]$
and, as long as the corresponding Racah matrices are built in the next papers
of the present series, for more sophisticated representations $R$.
In this kind of formulas, the knots/links are parameterized by braid words,
where numbers are substituted by the finger data (which are the braid words themselves).
Depending on relevance/need of the "propagators", the parametrization can become
even more interesting.

Of course, what is enumerated in this way, are {\it not} the {\it primary} knots and links:
there are many composite knots/links and the same knot/link appears many times.
New kind of questions appear, like the abundance of a given knot in a given sub-family,
and the search of the most "adequate" constituents (fingers?) for a given knot.
This can finally help to make the RT formalism really {\it local} so that it would
represent complicated primary knots as being composed from the "elementary blocks",
and finally find a classification based on the complexity of these blocks,
while at the present stage we can distinguish "complicated" knots only visually
and intuitively.
This is a long way to go, still we feel that time is coming to proceed to this kind
of analysis, and the data in \cite{ruskatlas} will be grouped to allow such attempts.

\section{On the highest weight calculus of ref.\cite{MMMknots2}}

The Racah matrix $U$ is a transformation matrix from one orthonormal basis (I) to another (II), which are defined as follows:
\be\label{2basis}
\left( R \otimes R \right) \otimes R \xrightarrow{ \ I \ } Q \ \ \xrightarrow{\ \ \ {\cal U} \ \ \ } \ \  R \otimes \left( R \otimes R \right) \xrightarrow{ \ II \ } Q
\ee
In our case $R=[2,1]$ and $Q$ is arbitrary, but most of them are equal to zero except for finitely many terms. To find nonzero terms, one can use the Littlewood-Richardson rule:
\be
\chi_{R_1} \cdot \chi_{R_2} &=& \sum_Q C_{R_1,R_2}^{Q} \cdot \chi_Q \\
\chi_{[2,1]}^3 &=& \chi_{[6,3]}+2\,\chi_{[6,2,1]}+\chi_{[6,1,1,1]}+2\,\chi_{[5,4]}+6\,\chi_{[5,3,1]}+4\,\chi_{[5,2,2]}+6\,\chi_{[5,2,1,1]}+ 2\,\chi_{[5,1,1,1,1]}+4\,\chi_{[4,4,1]} +   \nn \\ &+& 8\,\chi_{[4,3,2]} + 9\,\chi_{[4,3,1,1]}+9\,\chi_{[4,2,2,1]}+6\,\chi_{[4,2,1,1,1]}+\chi_{[4,1,1,1,1,1]}+2\,\chi_{[3,3,3]}+8\,\chi_{[3,3,2,1]}+\nn \\ &+& 4\,\chi_{[3,3,1,1,1]}+4\,\chi_{[3,2,2,2]} + 6\,\chi_{[3,2,2,1,1]} +2\,\chi_{[3,2,1,1,1,1]}+2\,\chi_{[2,2,2,2,1]}+\chi_{[2,2,2,1,1,1]}
\label{21dec}
\ee
Here $\chi_R$ is the character of the irreducible representation, which is the Schur function in the case of $SU(N)$, while $R$'s in this case are labelled by the Young diagrams. Hence, from now on, we identify the representations with the Young diagrams. In fact, we need the representation theory of $SU_q(N)$, but it is basically the same at $q$ non-equal to a root of unity.

The coefficients $C_{R_1,R_2}^{Q}$ count how many times the irreducible representation $Q$ appears in the decomposition, therefore they determine the size of the corresponding Racah matrix. Decomposition (\ref{21dec}) shows us that there are two matrices of the size $9\times9$, two matrices of the size $8\times8$, four matrices of the size $6\times6$, four matrices of the size $4\times4$, six matrices of the size $2\times2$ and four trivial "matrices" of the size $1\times1$.

We calculate the Racah matrix by definition, i.e. as a transformation matrix from the orthonormal basis (I) to the orthonormal basis (II). To this end, we construct the highest weight vectors in the basis (I) for each representation $Q$ from (\ref{21dec}) and same in the basis (II). To proceed, we need manifestly describe the action of lowering and raising operators $T_k^{\pm}$ on representations of $SU_q(N)$.

To this end, we use the Schur-Weyl duality and, first, realize the representation of $U_q(sl_N)$ in the space of tensors. For each representation labeled by Young diagram $Y = \{ Y_1\geq Y_2\geq\ldots\geq Y_l>0 \}$, we associate the following tensor with all possible permutations of indices:
\be\label{elY}
V_{i_1,\dots,i_{Y_1},j_1,\dots,j_{Y_2},k_1,\dots,k_{Y_3},\ldots}, \text{where} \\
i_1=\dots=i_{Y_1}=0, \ j_1=\dots=j_{Y_2}=0, \ k_1=\dots=k_{Y_3}=2, \ldots,
\ee
in other words, the number of zeros is equal to $Y_1$, the number of units is equal to $Y_2$, the number of deuces is equal to $Y_3$ and so on. Thus, every vector of the representation $Y$ can be written as a linear combination of elements (\ref{elY}).

Second, let us define the action of lowering and raising operators $T_k^{\pm}$. It is clear that for 1-tensors they act as follows:
\be
T_k^{+}: V_i \longrightarrow V_{i+1} \nn \\
T_k^{-}: V_i \longrightarrow V_{i-1}.
\ee
To extend this action to higher rank tensors, one needs a uniquely defined comultiplication $\Delta$ on $SU_q(N)$:
\be
\Delta(E_i) &=& 1\otimes E_i + E_i\otimes q^{H_\alpha}, \\
\Delta(F_i) &=& F_i\otimes 1 + q^{-H_\alpha}\otimes F_i, \\
\Delta(q^{H_\alpha}) &=& q^{H_\alpha}\otimes q^{H_\alpha},
\ee
where $E_i, F_i, q^{H_\alpha}$ are generators of $SU_q(N)$. Then, for 2-tensors one gets
\be
T_k^{+}: V_{i,j} \longrightarrow V_{i,j+1} + q^{H_k} V_{i+1,j} \nn \\
T_k^{-}: V_i \longrightarrow q^{-H_k} V_{i,j-1} + V_{i-1,j}.
\ee
Since $\Delta$ is co-associative, it is easy to extend $T_k^{\pm}$ actions to any rank tensors.

Now we are ready to construct highest weight vectors. Let us start with $R=[2,1]$ emerging in the decomposition $[2,1] \in [1]^{\otimes 3}$:
\be\label{hv21}
\left( [1] \otimes [1] \right) \otimes [1] = \left( [2] + [1,1] \right) \otimes [1] = \left( [3] + [2,1] \right) + \left( [2,1] + [1,1,1] \right).
\ee
These decompositions are also can be found by the Littlewood-Richardson rule, and the order of brackets $\left( [1] \otimes [1] \right) \otimes [1]$ or $[1] \otimes \left(  [1]  \otimes [1] \right)$ is not important here, just we should fix some particular order for this case once and forever. It is also not important which particular representation $[2,1]$ in (\ref{hv21}) we choose: the one which comes from $[2]\otimes[1]$ or from $[1,1]\otimes[1]$. We prefer  $[2,1] \in [1,1]\otimes[1]$, because our calculations are slightly simpler in this case. Now let us construct the highest weight vector for this representation.

It is clear that the highest weight vector of the fundamental representation $R=[1]$ is just $V_0$, because its indices correspond to $R=[1]$ (i.e. $\#0=1, \ \#i=0 \ \forall i>1$) and any $T_k^-$ cancels it. In the same way, it is clear that $V_{0,0}$ is the highest weight vector for the symmetric representation $R=[2]$. Now with the help of this vector let us construct the highest weight vector of $R=[1,1]$. First, one takes tensor product of the two fundamental representations and uses the raising operators to create the corresponding representation:
\be
T_1^+\left( V_0 \right) \otimes V_0 = V_1\otimes V_0 = V_{1,0}, \nn \\
V_0 \otimes T_1^+\left( V_0 \right) = V_0\otimes V_1 = V_{0,1}.
\ee
It is possible to create only two different vectors corresponding to $R=[1,1]$. They form a 2-dimensional vector space with coefficients in $\mathbb{Z}[q,q^{-1}]$. The highest weight vector lies in this space. To determine it, one requires that all lowering operators $T_k^-$ cancel on arbitrary vector from this space:
\be\label{hv11}
T_k^-\left( c_1 V_{1,0} + c_2 V_{0,1} \right) = 0 \ \  \Rightarrow \ \ c_1 =c, \ c_2=-q\cdot c,
\ee
where $c$ is an arbitrary constant. By definition, the Racah matrix is a transformation matrix from one orthonormal basis to another one, hence, all the highest weight vectors have to be unit vectors. This constraint determines $c$ uniquely
\be
c = \dfrac{1}{\sqrt{1+q^2}}.
\ee
Thus, we are done already, because $[1,1]\otimes [1]$ corresponds to $[2,1]$ itself and is canceled by all $T_k^-$:
\be
\dfrac{1}{\sqrt{1+q^2}} \cdot \left( V_{1,0} - q V_{0,1} \right) \otimes V_0 = \dfrac{1}{\sqrt{1+q^2}} \cdot \left( V_{1,0,0} - q V_{0,1,0} \right).
\ee

\bigskip

The described procedure allows us to construct the highest weight vector of any representation $R$ step by step. In particular, we found this way the highest weight vectors of all representations $Q$ in both bases (\ref{2basis}). These vectors are available at \cite{ruskatlas}, here let us give one simple example.

\paragraph{Example of $Q=[6,2,1]$.}
This representation comes from $[4,2]\otimes [2,1]$ and $[4,1,1]\otimes [2,1]$. The corresponding highest weight vectors in the (I) basis are
{\footnotesize
\be
[4,2]: \nn \\ \Big(-qV_{1,0,0,2,0,0,0,1,0}-q^2V_{2,0,0,1,0,0,0,1,0}-qV_{0,1,0,2,0,0,1,0,0}-q^2V_{0,2,0,1,0,0,1,0,0}-qV_{1,0,0,0,2,0,1,0,0}-q^2V_{2,0,0,0,1,0,1,0,0} \nn \\-q^4V_{0,2,0,0,1,0,0,1,0}+q^2V_{0,1,0,2,0,0,0,1,0}+q^3V_{0,2,0,1,0,0,0,1,0}+q^2V_{1,0,0,0,2,0,0,1,0}+q^3V_{2,0,0,0,1,0,0,1,0}+q^4V_{0,1,0,0,1,0,0,2,0}\nn \\-q^3V_{0,1,0,1,0,0,0,2,0}-q^3V_{1,0,0,0,1,0,0,2,0}+q^2V_{1,0,0,1,0,0,0,2,0}-q^3V_{0,1,0,0,1,0,2,0,0}+q^2V_{0,1,0,1,0,0,2,0,0}+q^2V_{1,0,0,0,1,0,2,0,0}\nn \\-qV_{1,0,0,1,0,0,2,0,0}+q^6V_{0,1,0,0,1,0,0,2,0}-q^5V_{0,1,0,1,0,0,0,2,0}-q^5V_{1,0,0,0,1,0,0,2,0}+q^4V_{1,0,0,1,0,0,0,2,0}-q^5V_{0,1,0,0,1,0,2,0,0}\nn \\ +q^4V_{0,1,0,1,0,0,2,0,0}+q^4V_{1,0,0,0,1,0,2,0,0}-q^3V_{1,0,0,1,0,0,2,0,0}+q^2V_{0,1,0,0,2,0,1,0,0}+q^3V_{0,2,0,0,1,0,1,0,0}+V_{1,0,0,2,0,0,1,0,0}\nn \\ +qV_{2,0,0,1,0,0,1,0,0}-q^3V_{0,1,0,0,2,0,0,1,0} \Big)\cdot \dfrac{1}{\sqrt {{q}^{4}+{q}^{2}+1} \left( {q}^{2}+1 \right) ^{2}};\nn \\
\nn \\
\text{[}4,1,1\text{]}: \nn \\
\Big( q^4V_{0,1,0,0,2,0,0,1,0}-q^3V_{0,1,0,2,0,0,0,1,0}-q^3V_{0,2,0,0,1,0,0,1,0}+q^2V_{0,2,0,1,0,0,0,1,0}-q^3V_{1,0,0,0,2,0,0,1,0}+q^2V_{1,0,0,2,0,0,0,1,0} \nn \\ +q^2V_{2,0,0,0,1,0,0,1,0}-qV_{2,0,0,1,0,0,0,1,0}-q^3V_{0,1,0,0,2,0,1,0,0}+q^2V_{0,1,0,2,0,0,1,0,0}+q^2V_{0,2,0,0,1,0,1,0,0}-qV_{0,2,0,1,0,0,1,0,0}\nn \\ +q^2V_{1,0,0,0,2,0,1,0,0}-qV_{1,0,0,2,0,0,1,0,0}-qV_{2,0,0,0,1,0,1,0,0}+V_{2,0,0,1,0,0,1,0,0} \Big)\cdot \dfrac{1}{\left( {q}^{
2}+1 \right) ^{2}} \nn
\ee
}
and in the (II) basis
{\footnotesize
\be
[4,2]: \nn \\
\Big( -q^3V_{0,2,0,0,1,0,0,1,0}+q^2V_{0,2,0,0,1,0,1,0,0}+q^2V_{0,2,0,1,0,0,0,1,0}+q^2V_{2,0,0,0,1,0,0,1,0}-q^5V_{0,2,0,0,1,0,0,1,0}+q^4V_{2,0,0,0,1,0,0,1,0} \nn \\ +q^4V_{0,2,0,0,1,0,1,0,0}-q^3V_{2,0,0,0,1,0,1,0,0}+q^4V_{0,2,0,1,0,0,0,1,0}-q^3V_{2,0,0,1,0,0,0,1,0}-q^3V_{0,2,0,1,0,0,1,0,0}+q^2V_{2,0,0,1,0,0,1,0,0} \nn \\ +q^5V_{0,1,0,0,1,0,0,2,0}-q^4V_{1,0,0,0,1,0,0,2,0}-q^4V_{0,1,0,0,1,0,2,0,0}+q^3V_{1,0,0,0,1,0,2,0,0}+q^6V_{0,1,0,0,2,0,0,1,0}-q^5V_{1,0,0,0,2,0,0,1,0} \nn \\ -q^5V_{0,1,0,0,2,0,1,0,0}+q^4V_{1,0,0,0,2,0,1,0,0}-q^4V_{0,1,0,1,0,0,0,2,0}+q^3V_{1,0,0,1,0,0,0,2,0}+q^3V_{0,1,0,1,0,0,2,0,0}-q^2V_{1,0,0,1,0,0,2,0,0} \nn \\ -q^5V_{0,1,0,2,0,0,0,1,0}+q^4V_{1,0,0,2,0,0,0,1,0}+q^4V_{0,1,0,2,0,0,1,0,0}-q^3V_{1,0,0,2,0,0,1,0,0}-qV_{2,0,0,1,0,0,0,1,0}-qV_{0,2,0,1,0,0,1,0,0} \nn \\ -qV_{2,0,0,0,1,0,1,0,0}+V_{2,0,0,1,0,0,1,0,0} \Big) \cdot \dfrac{1}{\sqrt {{q}^{4}+{q}^{2}+1} \left( {q}^{2}+1 \right) ^{2}}; \nn \\
\nn
\ee
\be
\text{[}4,1,1\text{]}: \nn \\
\Big( q^4V_{0,1,0,0,1,0,0,2,0}-q^3V_{1,0,0,0,1,0,0,2,0}-q^3V_{0,1,0,0,1,0,2,0,0}+q^2V_{1,0,0,0,1,0,2,0,0}-q^3V_{0,1,0,0,2,0,0,1,0}+q^2V_{1,0,0,0,2,0,0,1,0} \nn \\ +q^2V_{0,1,0,0,2,0,1,0,0}-qV_{1,0,0,0,2,0,1,0,0}-q^3V_{0,1,0,1,0,0,0,2,0}+q^2V_{1,0,0,1,0,0,0,2,0}+q^2V_{0,1,0,1,0,0,2,0,0}-qV_{1,0,0,1,0,0,2,0,0} \nn \\ +q^2V_{0,1,0,2,0,0,0,1,0}-qV_{1,0,0,2,0,0,0,1,0}-qV_{0,1,0,2,0,0,1,0,0}+V_{1,0,0,2,0,0,1,0,0} \Big) \cdot \dfrac{1}{\left( {q}^{2}+1 \right) ^{2}}. \nn
\ee
}
From their explicit forms we find the following rotation matrix, which is Racah matrix by definition
\be
{\cal U}_{[6,2,1]} =   \left(\begin{array}{cc} \frac{1}{[2]} & -\frac{\sqrt{[3]}}{[2]} \\ \\
-\frac{\sqrt{[3]}}{[2]} & -\frac{1}{[2]}
\end{array}\right)
\ee

\bigskip

\paragraph{Example of $Q=[3,2,1]$.} Finally let us discuss the case of multiplicities. Indeed, in the product of two $[2,1]$ there are two representations $[3,2,1]$:
\be
\chi_{[2,1]}^2=\chi_{[4,2]}+\chi_{[4,1,1]}+\chi_{[3,3]}+2\chi_{[3,2,1]}+\chi_{[3,1,1,1]}+\chi_{[2,2,2]}+\chi_{[2,2,1,1]}.
\ee
In the calculation of the highest weight vector, one obtains the two-dimensional vector space instead of one-dimensional space like in (\ref{hv11}). How to choose the two highest weight vectors corresponding to two different $[3,2,1]$ representations? We cannot choose them arbitrarily, because the ${\cal R}$-matrix is diagonal only in a particular basis. It turns out that it is enough to put one more condition to determine two different highest weight vectors of $[3,2,1]$ corresponding to the diagonal ${\cal R}$-matrix. It is the following condition: one vector has to belong to the \textit{symmetric} tensor square product $[2,1]^{\otimes 2}$, while the other one belongs to the \textit{antisymmetric} tensor square product. Indeed, when one decomposes the tensor square of the representation $\lambda$ into irreducible representations $\mu_i$, every representation $\mu_i$ comes either from the symmetric or antisymmetric squares. In terms of the highest weight vectors, it means that the vectors are either symmetric or antisymmetric under the following indices permutation:
\be
V_{i_1,...,i_{|R|},i_{|R|+1},...,i_{2|R|}} \ \longrightarrow \ V_{i_{|R|+1},...,i_{2|R|},i_1,...,i_{|R|}},
\ee
where $|R| = \sum_iR_i$. Thus, all irreducible representations of the tensor square of $[2,1]$ can be divided into two groups, symmetric and antisymmetric (underlined):
\be
\l[2,1]\otimes [2,1] = [4,2]+\underline{[4,1,1]}+\underline{[3,3]}+[3,2,1]+\underline{[3,2,1]}+{[3,1,1,1]}+[2,2,2]+\underline{[2,2,1,1]}.
\ee

\bigskip

This completes the tool box for our calculations. At the end of this section we list two highest weight vectors of $[3,2,1]$, symmetric and antisymmetric:
{\footnotesize
\be [3,2,1]: \nn \\
{q}^{3}V_{1,0,0,2,0,1}-{q}^{5}V_{0,1,0,2,1,0}+{q}^{4}V_{1,0,0,2,1,0}-{q}^{3}V_{0,1,1,0,2,0}+{q}^{2}V_{1,0,1,0,2,0}+{q}^{2}V_{0,1,1,2,0,0}-qV_{1,0,1,2,0,0} \nn \\ -{q}^{5}V_{0,1,1,0,2,0}+{q}^{4}V_{1,0,1,0,2,0}+{q}^{4}V_{0,1,1,2,0,0}-{q}^{3}V_{1,0,1,2,0,0}-{q}^{4}V_{0,2,0,0,1,1}+{q}^{3}V_{2,0,0,0,1,1}+{q}^{3}V_{0,2,0,1,0,1} \nn \\ -{q}^{4}V_{0,1,1,0,2,0}+{q}^{3}V_{1,0,1,0,2,0}+{q}^{3}V_{0,1,1,2,0,0}-{q}^{2}V_{1,0,1,2,0,0}-{q}^{3}V_{0,2,0,0,1,1}+{q}^{2}V_{2,0,0,0,1,1}+{q}^{2}V_{0,2,0,1,0,1} \nn \\ -qV_{2,0,0,1,0,1}+{q}^{5}V_{0,1,0,0,1,2}-{q}^{4}V_{1,0,0,0,1,2}+{q}^{6}V_{0,1,0,0,2,1} -{q}^{5}V_{1,0,0,0,2,1} -{q}^{4}V_{0,1,0,1,0,2}+{q}^{3}V_{1,0,0,1,0,2} \nn \\ -{q}^{5}V_{0,1,0,2,0,1} +{q}^{4}V_{1,0,0,2,0,1}+{q}^{2}V_{0,2,1,0,1,0}+{q}^{3}V_{1,2,0,0,1,0}-qV_{2,0,1,0,1,0}-{q}^{2}V_{2,1,0,0,1,0}-qV_{0,2,1,1,0,0} \nn \\ -{q}^{2}V_{1,2,0,1,0,0}+qV_{2,1,0,1,0,0}-{q}^{5}V_{0,2,0,0,1,1}+{q}^{4}V_{2,0,0,0,1,1}+{q}^{4}V_{0,2,0,1,0,1}-{q}^{3}V_{2,0,0,1,0,1}-{q}^{2}V_{2,0,0,1,0,1} \nn \\ +{q}^{2}V_{0,1,2,0,1,0}+{q}^{3}V_{0,2,1,0,1,0}-qV_{0,1,2,1,0,0}-{q}^{2}V_{0,2,1,1,0,0}-qV_{1,0,2,0,1,0}-{q}^{2}V_{2,0,1,0,1,0}+qV_{2,0,1,1,0,0} \nn \\ +{q}^{5}V_{0,1,0,0,2,1}-{q}^{4}V_{1,0,0,0,2,1}+{q}^{6}V_{0,1,0,1,2,0}-{q}^{5}V_{1,0,0,1,2,0}-{q}^{4}V_{0,1,0,2,0,1}+V_{2,0,1,1,0,0}+V_{1,0,2,1,0,0} \nn
\ee

\be \underline{[3,2,1]}: \nn \\
-{q}^{3}V_{1,0,0,2,0,1}+{q}^{5}V_{0,1,0,2,1,0}-{q}^{4}V_{1,0,0,2,1,0}+{q}^{3}V_{0,1,1,0,2,0}-{q}^{2}V_{1,0,1,0,2,0}-{q}^{2}V_{0,1,1,2,0,0}   +qV_{1,0,1,2,0,0} \nn \\ +{q}^{5}V_{0,1,1,0,2,0}-{q}^{4}V_{1,0,1,0,2,0}-{q}^{4}V_{0,1,1,2,0,0}+{q}^{3}V_{1,0,1,2,0,0}+{q}^{4}V_{0,2,0,0,1,1}  -{q}^{3}V_{2,0,0,0,1,1}-{q}^{3}V_{0,2,0,1,0,1} \nn \\ -{q}^{4}V_{0,1,1,0,2,0}+{q}^{3}V_{1,0,1,0,2,0}+{q}^{3}V_{0,1,1,2,0,0}-{q}^{2}V_{1,0,1,2,0,0}  -{q}^{3}V_{0,2,0,0,1,1}+{q}^{2}V_{2,0,0,0,1,1}+{q}^{2}V_{0,2,0,1,0,1} \nn \\ -qV_{2,0,0,1,0,1}+{q}^{5}V_{0,1,0,0,1,2}-{q}^{4}V_{1,0,0,0,1,2}    +{q}^{6}V_{0,1,0,0,2,1}- {q}^{5}V_{1,0,0,0,2,1}-{q}^{4}V_{0,1,0,1,0,2}+{q}^{3}V_{1,0,0,1,0,2} \nn \\ -{q}^{5}V_{0,1,0,2,0,1} +{q}^{4}V_{1,0,0,2,0,1}   +{q}^{2}V_{0,2,1,0,1,0}+{q}^{3}V_{1,2,0,0,1,0}-qV_{2,0,1,0,1,0}-{q}^{2}V_{2,1,0,0,1,0}-qV_{0,2,1,1,0,0} \nn \\ -{q}^{2}V_{1,2,0,1,0,0}    +qV_{2,1,0,1,0,0}-{q}^{5}V_{0,2,0,0,1,1}+{q}^{4}V_{2,0,0,0,1,1}+{q}^{4}V_{0,2,0,1,0,1}-{q}^{3}V_{2,0,0,1,0,1}+{q}^{2}V_{2,0,0,1,0,1}  \nn \\  -{q}^{2}V_{0,1,2,0,1,0}-{q}^{3}V_{0,2,1,0,1,0}+qV_{0,1,2,1,0,0}+{q}^{2}V_{0,2,1,1,0,0}+qV_{1,0,2,0,1,0}+{q}^{2}V_{2,0,1,0,1,0}  -qV_{2,0,1,1,0,0} \nn \\ -{q}^{5}V_{0,1,0,0,2,1}+{q}^{4}V_{1,0,0,0,2,1}-{q}^{6}V_{0,1,0,1,2,0}+{q}^{5}V_{1,0,0,1,2,0}+{q}^{4}V_{0,1,0,2,0,1}  +V_{2,0,1,1,0,0}-V_{1,0,2,1,0,0} \nn
\ee
}


\section{Non-symmetric representation $R=[2,1]$}

In this section we construct the ${\cal R}$-matrices and the mixing matrices for the non-symmetric representation $R=[2,1]$ for the three strand case in order to evaluate the HOMFLY polynomials in representation $R=[2,1]$. The polynomials themselves are discussed in further sections. Since our approach is the group theoretical one, the ${\cal R}$-matrices are obtained in the vertical framing.

\subsection{Two strands}

Since
$\chi_{[2,1]}^2=\chi_{[4,2]}+\chi_{[4,1,1]}+\chi_{[3,3]}+2\chi_{[3,2,1]}+\chi_{[3,1,1,1]}+\chi_{[2,2,2]}+\chi_{[2,2,1,1]}$,
\be\label{212}
\l[2,1]\otimes [2,1] = [4,2]+\underline{[4,1,1]}+\underline{[3,3]}+[3,2,1]+\underline{[3,2,1]}
+{[3,1,1,1]}+[2,2,2]+\underline{[2,2,1,1]}
\ee
Underlined are representations from antisymmetric square, for which a minus sign should be
added to the eigenvalues of the ${\cal R}$-matrices\footnote{One can determine if the representation belongs to symmetric or antisymmetric square looking at the
symmetry of its highest weight vector as it was done in the previous section. However, technically the simplest way to find a decomposition of square of the
representation $R$ into symmetric and antisymmetric parts is to use the plethysm: one has to expand
\be
{1\over 2}\chi_R(p)^2\pm {1\over 2}\chi_R(p^{(2)})=\sum_{Q_{\pm}} \chi_{Q_{\pm}}(p)
\ee
in order to find $Q_{\pm}$'s that emerge in the symmetric (+) and antisymmetric (-) squares. Here $|Q_{\pm}|=2|R|$,
$\chi_R(p)$ is the Schur function of variables
$p_k=\Tr g^k$, and the Adams 2-transformation is given by $p_k\to p_{2k}\equiv p^{(2)}_k$.
}.
In particular, the two $[3,2,1]$ come from symmetric and antisymmetric squares and thus are
well defined basis vectors in the two-dimensional space as it was explained at the end of the previous section.
The  Casimir (cut-and-join) eigenvalues $\varkappa_Q = \sum_{(i,j)\in Q} (i-j)$
are $\varkappa_{[4,2]}=5$, $\varkappa_{[4,1,1]}=3$, $\varkappa_{[3,3]}=3$,
$\varkappa_{[3,2,1]}=0$, $\varkappa_{[3,1,1,1]}=-3$, $\varkappa_{[2,2,2]}=-3$, $\varkappa_{[2,2,1,1]}=-5$,
therefore, the Rosso-Jones formula \cite{RJ} for the two-strand torus link $T[2,n]$ gives:
\be
H_{[2,1]\otimes[2,1]}^{[2,2n]} =
q^{10n}\chi^*_{[4,2]}+q^{6n}\chi^*_{[4,1,1]}+q^{6n}\chi^*_{[3,3]}+2\chi^*_{[3,2,1]}+q^{-6n}\chi^*_{[3,1,1,1]}+
q^{-6n}\chi^*_{[2,2,2]}+q^{-10n}\chi^*_{[2,2,1,1]}
\ee
where the asterisk means that the character is calculated at "the topological locus" $p_k={A^k-A^{-k}\over q^k-q^{-k}}$.

For two-strand torus knot $T[2,2n+1]$ one needs slightly more complicated Adams coefficients
$\widehat{\rm Ad}_2 \chi_{[2,1]} =\chi_{[4,2]}-\chi_{[4,1,1]}-\chi_{[3,3]}
  +\chi_{[3,1,1,1]}+\chi_{[2,2,2]}-\chi_{[2,2,1,1]}
$
and
\be
H_{[2,1] }^{[2,2n+1]} =
q^{5(2n+1)}\chi^*_{[4,2]}-q^{3(2n+1)}\chi^*_{[4,1,1]}-q^{3(2n+1)}\chi^*_{[3,3]}
+q^{-3(2n+1)}\chi^*_{[3,1,1,1]}+q^{-3(2n+1)}\chi^*_{[2,2,2]}-q^{-5(2n+1)}\chi^*_{[2,2,1,1]}
\ee
Note that $\chi_{[3,2,1]}$ does not contribute in this case, which means that the two eigenvalues
have different signs.

\subsection{Three strands}

In the three-strand case, one needs
{\footnotesize
\be
\l[2,1]^{\otimes 3} = \Big([4,2]+[4,1,1]+[3,3]+2\cdot[3,2,1]+[3,1,1,1]+[2,2,2]+[2,2,1,1]\Big)\otimes [2,1] =
\nn
\ee
\be
= \Big([6,3]+[6,2,1]+[5,4]+2\cdot [5,3,1]+[5,2,2]+[5,2,1,1]+[4,4,1]+[4,3,2]+[4,3,1,1]+[4,2,2,1]\Big) + \nn \\
+ \Big([6,2,1]+[6,1,1,1]+[5,3,1]+[5,2,2]+ 2\cdot[5,2,1,1]+[5,1,1,1,1]+  [4,3,2]+[4,3,1,1]+[4,2,2,1]+[4,2,1,1,1]\Big) + \nn\\
+ \Big(   [5,4]+[5,3,1]+[4,4,1]+[4,3,2]+[4,3,1,1]+[3,3,2,1]\Big)+ \nn \\
+2\cdot\Big([5,3,1]+[5,2,2]+[5,2,1,1]+[4,4,1]+ 2\cdot[4,3,2] + 2\cdot[4,3,1,1]+2\cdot[4,2,2,1]+\nn\\+[4,2,1,1,1]
+[3,3,3]+2\cdot[3,3,2,1]+[3,3,1,1,1]+[3,2,2,2]+[3,2,2,1,1]\Big)+\nn \\
+\Big([5,2,1,1]+[5,1,1,1,1]+[4,3,1,1]+[4,2,2,1]+2\cdot[4,2,1,1,1]+[4,1,1,1,1,1]+[3,3,2,1]+[3,3,1,1,1]+[3,2,2,1,1]+[3,2,1,1,1,1]\Big)+\nn\\
+\Big([4,3,2]+[4,2,2,1]+[3,3,2,1]+[3,2,2,2]+[3,2,2,1,1]+[2,2,2,2,1]\Big)+\nn \\
+\Big([4,3,1,1]+[4,2,2,1]+[4,2,1,1,1]+[3,3,2,1]+[3,3,1,1,1]+[3,2,2,2]+2\cdot [3,2,2,1,1]+[3,2,1,1,1,1]+[2,2,2,2,1]+[2,2,2,1,1,1]\Big) =
\nn\\ \nn \\ \nn
\ee
\be
= \underbrace{[6,3]}_{[4,2]}+\underbrace{2\cdot[6,2,1]}_{[4,2]+[4,1,1]}+\underbrace{[6,1,1,1]}_{[4,1,1]}+
\underbrace{2\cdot[5,4]}_{[4,2]+[3,3]}+\underbrace{6\cdot [5,3,1]}_{2\cdot[4,2]+[4,1,1]+[3,3]+2\cdot[3,2,1]}
+\underbrace{4\cdot[5,2,2]}_{[4,2]+[4,1,1]+2\cdot[3,2,1]}
+\underbrace{6\cdot[5,2,1,1]}_{[4,2]+2\cdot[4,1,1]+2\cdot[3,2,1]+[3,1,1,1]}
+\nn\\
+\underbrace{2\cdot[5,1,1,1,1]}_{[4,1,1]+[3,1,1,1]}
+\underbrace{4\cdot[4,4,1]}_{[4,2]+[3,3]+2\cdot[3,2,1]}
+\underbrace{8\cdot[4,3,2]}_{[4,2]+[4,1,1]+[3,3]+4\cdot[3,2,1]+[2,2,2]}
+\underbrace{9\cdot[4,3,1,1]}_{[4,2]+[4,1,1]+[3,3]+4\cdot[3,2,1]+[3,1,1,1]+[2,2,1,1]}+\nn\\
+\underbrace{9\cdot[4,2,2,1]}_{[4,2]+[4,1,1]+4\cdot[3,2,1]+[3,1,1,1]+[2,2,2]+[2,2,1,1]}
+\underbrace{6\cdot[4,2,1,1,1]}_{[4,1,1]+2\cdot[3,2,1]+2\cdot[3,1,1,1]+[2,2,1,1]}+\underbrace{[411111]}_{[3111]}+\nn\\
+\underbrace{2\cdot[3,3,3]}_{2\cdot[3,2,1]} + \underbrace{8\cdot[3,3,2,1]}_{[3,3]+4\cdot[3,2,1]+[3,1,1,1]+[2,2,2]+[2,2,1,1]}
+\underbrace{4\cdot[3,3,1,1,1]}_{2\cdot[3,2,1]+[3,1,1,1]+[2,2,1,1]}
+\underbrace{4\cdot[3,2,2,2]}_{2\cdot[3,2,1]+[2,2,2]+[2,2,1,1]}+\nn\\
+\underbrace{6\cdot[3,2,2,1,1]}_{2\cdot[3,2,1]+[3,1,1,1]+[2,2]+2\cdot[2,2,1,1]}
+ \underbrace{2\cdot[3,2,1,1,1,1]}_{[3,1,1,1]+[2,2,1,1]}+\underbrace{2\cdot[2,2,2,2,1]}_{[2,2]+[2,2,1,1]}
+\underbrace{[2,2,2,1,1,1]}_{[2,2,1,1]}
\nn
\ee}
For the three-strand torus links one has ($n=3k\pm 1$):
\be
H_{[2,1]}^{[3,n]} =
q^{-10n}\chi_{[6,3]}-q^{-8n}\chi^*_{[6,2,1]}+q^{-6n}\chi^*_{[6,1,1,1]}
-q^{-8n}\chi^*_{[5,4]}+q^{-4n}\chi^*_{[5,2,2]}-\chi^*_{[5,1,1,1,1]}
+q^{-4n}\chi^*_{[4,4,1]} -q^{-2n}\chi^*_{[4,3,2]} + \nn\\
\!\!\!\!\!\!\!\!\!\!\!\! +q^{6n}\chi^*_{[4,1,1,1,1,1]}
+2\chi^*_{[3,3,3]}-q^{2n}\chi^*_{[3,3,2,1]}+q^{4n}\chi^*_{[3,3,1,1,1]}+q^{4n}\chi^*_{[3,2,2,2]}-q^{8n}\chi^*_{[3,2,1,1,1,1]}
-q^{8n}\chi^*_{[2,2,2,2,1]}+q^{10n}\chi^*_{[2,2,2,1,1,1]}
\nn\ee

\subsection{Mixing matrices\label{eigenvhyp}}

The $1\times 1$ ${\cal R}$-matrices are
\be
{\cal R}_{[6,3]} = q^{-\varkappa_{[4,2]}} = q^{-5}
\ee
and
\be
{\cal R}_{[6,1,1]} = -q^{\varkappa_{[4,1,1]}} = -q^{-3}
\ee

As for the higher size mixing matrices, one can calculate them using the results of the previous section. However, for the sizes of matrices up to five there is the eigenvalue conjecture \cite{eigen} that allows one to obtain the mixing matrices immediately. That is,
according to the eigenvalue hypothesis of \cite{eigen} (see also \cite{mathmix})
the $2\times 2$ and $4\times 4$ mixing matrices are expressed through the normalized eigenvalues of ${\cal R}$-matrix
$\tilde\xi_Q$:
\be
\tilde \xi_i = \frac{\xi_i}{\left(\pm\prod_{j=1}^p \xi_j \right)^{1/p}}
\ee
where the product runs over all $Q_j \in R\otimes R$, $j=1\ldots p$.
For $2\times 2$ mixing matrices, \cite{eigen} conjectures:
\be
U_{ii} = \frac{\sqrt{-1}}{\prod_{k\neq i}\tilde\xi_{ik}},
\ \ \ \ \ \ \ \ \ \ \
U_{ij} =\sqrt{\frac{(\tilde\xi_1^2-1+\tilde\xi_1^{-2})}{\tilde\xi_{ij}^2\tilde\xi_{ik}\tilde\xi_{jk}}}=
\sqrt{\frac{(\tilde\xi_2^2-1+\tilde\xi_2^{-2})}{\tilde\xi_{ij}^2\tilde\xi_{ik}\tilde\xi_{jk}}}
\ee
while for $4\times 4$ mixing matrices, \cite{eigen} conjectures:
\be
U_{ii} = \frac{\tilde\xi_i^2 \sum_{(k<l)\neq i}\tilde\xi_k\tilde\xi_l
- \tilde\xi_i \sum_{k\neq i}\tilde\xi_k}{\prod_{k\neq i}\tilde\xi_{ik}},
\ \ \ \ \ \ \ \ \ \ \
U_{ij} =\sqrt{\frac{(\tilde\xi_i^2-1)(\tilde\xi_j^2-1)}{\tilde\xi_{ij}^2}
\prod_{k\neq i,j}\left(\frac{\tilde\xi_i\tilde\xi_k-1+(\tilde\xi_i\tilde\xi_k)^{^{-1}}}
{\tilde\xi_{ik}\tilde\xi_{jk}}\right)}
\ee
(making use of $\prod_{k=1}^4\tilde \xi_k$ one can check that the last product is symmetric under the
permutation $i\leftrightarrow j$).

\bigskip

Using these formulas, one easily gets
{\footnotesize
\be
{\cal R}_{[6,2,1]} =
\left(\begin{array}{cc} q^{-\varkappa_{[4,2]}}&  \\ \\
& -q^{-\varkappa_{[4,1,1]}} \end{array}\right)
=
\left(\begin{array}{cc} \frac{1}{q^5}&        \\ \\ & -\frac{1}{q^3} \end{array}\right)
\ \ \ \ \ \ \ \
{\cal U}_{[6,2,1]} =   \left(\begin{array}{cc} \frac{1}{[2]} & -\frac{\sqrt{[3]}}{[2]} \\ \\
-\frac{\sqrt{[3]}}{[2]} & -\frac{1}{[2]}
\end{array}\right)
\ee

\be
{\cal R}_{[5,4]} =
\left(\begin{array}{cc} q^{-\varkappa_{[4,2]}}&  \\ \\
& -q^{-\varkappa_{[3,3]}} \end{array}\right)
=\left(\begin{array}{cc} \frac{1}{q^5}&   \\ \\ & -\frac{1}{q^3} \end{array}\right)
\ \ \ \ \ \ \ \ \
{\cal  U}_{[5,4]} =   \left(\begin{array}{cc} \frac{1}{[2]} & -\frac{\sqrt{[3]}}{[2]} \\ \\
-\frac{\sqrt{[3]}}{[2]} & -\frac{1}{[2]}
\end{array}\right)
\ee

\bigskip

\be
{\cal R}_{[5,1,1,1,1]} =
\left(\begin{array}{cc} -q^{-\varkappa_{[4,1,1]}}&  \\ \\
& q^{-\varkappa_{[3,1,1,1]}} \end{array}\right)
=
\left(\begin{array}{cc} \frac{1}{q^5}&   \\ \\ & -\frac{1}{q^3} \end{array}\right)
\ \ \ \ \ \ \ \ \ \ \
{\cal U}_{[5,1,1,1,1]} = \left(\begin{array}{cc} \frac{1}{q^3+q^{-3}}
& \frac{\sqrt{q^6+1+q^{-6}}}{q^3+q^{-3}} \\ \\
\frac{\sqrt{q^6+1+q^{-6}}}{q^3+q^{-3}} & -\frac{1}{q^3+q^{-3}} \end{array}\right)
\ee
\be
{\cal R}_{[3,3,3]} =
\left(\begin{array}{cc} q^{-\varkappa_{[3,2,1]}}&  \\ \\
& -q^{-\varkappa_{[3,2,1]}} \end{array}\right)
=
\left(\begin{array}{cc} 1&   \\ \\ & -1 \end{array}\right)
\ \ \ \ \ \ \ \ \ \
{\cal U}_{[3,3,3]} = \left(\begin{array}{cc} 1 &0 \\ \\ 0 & 1 \end{array}\right)
\ee

\bigskip

\be
{\cal R}_{[5,2,2]} = \left(\begin{array}{cccc}
q^{-\varkappa_{[4,2]}}&&& \\ \\
& -q^{-\varkappa_{[4,1,1]}}&& \\ \\
&& -q^{-\varkappa_{[3,2,1]}} & \\ \\
&&& q^{-\varkappa_{[3,2,1]}}
\end{array}\right)
= \left(\begin{array}{cccc}
\frac{1}{q^5} &&& \\ \\ & -\frac{1}{q^3} && \\ \\ &&-1& \\ \\ &&&1
\end{array}\right)
\ee

\be
U_{[5,2,2]}
=\left(\begin{array}{cccc}
-\frac{1}{[2][5]} & \frac{1}{[2]}\sqrt{\frac{[3]}{[5]}}
& -\beta & \alpha  \\ \\
\frac{1}{[2]}\sqrt{\frac{[3]}{[5]}} & \frac{1}{[2]}
&\delta&  \gamma  \\ \\
\beta & -\delta & \frac{[1/2]\cdot \{q\}^2}{2(q^{5/2}+q^{-5/2})} & \frac{[2]}{2\sqrt{[5]}} \\ \\
 \alpha & \gamma & -\frac{[2]}{2\sqrt{[5]}} & -\frac{q^{1/2}+q^{-1/2}}{2\cdot [5/2]}
\end{array}\right)
\ee
}
with
\be
\begin{array}{ll}
\alpha=
\sqrt{\frac{q^4+1+q^{-4}}{2}}
\frac{\sqrt{[3](q+1+q^{-1})}}{[5]}\sqrt{\frac{q^{5/2}+q^{-5/2}}{q^{3/2}+q^{-3/2}}} &
\beta=\sqrt{\frac{q^4+1+q^{-4}}{2}}
\sqrt{\frac{q-1+q^{-1}}{[5]}}\sqrt{\frac{q^{3/2}+q^{-3/2}}{q^{5/2}+q^{-5/2}}}\nn \\ \nn \\
\gamma=  \sqrt{\frac{q^4+1+q^{-4}}{2}}
\sqrt{\frac{q-1+q^{-1}}{[3][5]}}\sqrt{\frac{q^{5/2}+q^{-5/2}}{q^{3/2}+q^{-3/2}}}
& \delta =\sqrt{\frac{q^4+1+q^{-4}}{2}}
\frac{\sqrt{q+1+q^{-1}}}{[3]}\sqrt{\frac{q^{3/2}+q^{-3/2}}{q^{5/2}+q^{-5/2}}}
\end{array}
\ee
so that
\be\label{rot}
\gamma^2+\delta^2 = \frac{[6]}{[2][5]}, \ \ \ \gamma^2-\delta^2=\frac{[6]}{[3][5]}, \ \ \
2\gamma\delta = \frac{[6]}{[2][3]\sqrt{[5]}}
\ee

%

{\footnotesize
\be
{\cal R}_{[4,4,1]} = \left(\begin{array}{cccc}
q^{-\varkappa_{[4,2]}} &&& \\ \\
& -q^{-\varkappa_{[3,3]}}&& \\ \\
&& -q^{-\varkappa_{[3,2,1]}} & \\ \\
&&& q^{-\varkappa_{[3,2,1]}}
\end{array}\right)
= \left(\begin{array}{cccc}
\frac{1}{q^5} &&& \\ \\ & -\frac{1}{q^3} && \\ \\ &&-1& \\ \\ &&&1
\end{array}\right)
\ee
\be
{\cal U}_{[4,4,1]} = {\cal U}_{[5,2,2]}
%
\ee
}

For $6\times 6$, $8\times 8$ and $9\times 9$ matrices the eigenvalues of ${\cal R}$
do not define the mixing matrix uniquely, thus, the mixing matrices need to be calculated
by the direct method of \cite{MMMknots2}. The result is listed in Appendix A.

\subsection{Euler angles}

The mixing matrices usually have a simple parameterization in terms of elementary buildings block associated with simple
choices of Euler angles. For instance, using formulas (\ref{rot}), one has perform a rotation at angle $\theta$ with $c=\cos\theta = \frac{\delta}{\sqrt{\gamma^2+\delta^2}}$ and
$s=\sin\theta = \frac{\gamma}{\sqrt{\gamma^2+\delta^2}}$ in the $[3,2,1]$ sector
that converts the off-diagonal $2\times 2$ block of $U_{[5,2,2]}$ into:
\be
\begin{array}{ll}
\alpha s-\beta c =\frac{\alpha\gamma-\beta\delta}{\sqrt{\gamma^2+\delta^2}}
=\frac{1}{[5]}\sqrt{\frac{[6]}{[2][3]}}
& \alpha c+\beta s =\frac{\alpha\delta+\beta\gamma}{\sqrt{\gamma^2+\delta^2}}=\sqrt{\frac{[6][2]}{[3][5]}}\\
\gamma s+\delta c = \sqrt{\frac{[6]}{[2][5]}} & \gamma c -\delta s = 0
\\ \\
\alpha s+\beta c = \frac{\alpha\gamma+\beta\delta}{\sqrt{\gamma^2+\delta^2}}
=\frac{[4]}{[5]}\sqrt{\frac{[6]}{[2][3]}} &
\alpha c -\beta s = \frac{\alpha\delta-\beta\gamma}{\sqrt{\gamma^2+\delta^2}}=\sqrt{\frac{[6]}{[2][3][5]}}
\\
\gamma s-\delta c =\frac{\gamma^2-\delta^2}{\sqrt{\gamma^2+\delta^2}}=
\frac{1}{[3]}\sqrt{\frac{[6][2]}{[5]}}&
\gamma c+\delta s =\frac{2\gamma\delta}{\sqrt{\gamma^2+\delta^2}}=\frac{1}{[3]}\sqrt{\frac{[6]}{[2]}}
\end{array}
\ee

Thus, one gets the matrix
{\footnotesize
\be
\tilde U_{[5,2,2]} = \left(\begin{array}{cccc}
-\frac{1}{[2][5]} & \frac{1}{[2]}\sqrt{\frac{[3]}{[5]}} & \frac{1}{[5]}\sqrt{\frac{[6]}{[3][2]}} & \sqrt{\frac{[6][2]}{[5][3]}} \\ \\
\frac{1}{[2]}\sqrt{\frac{[3]}{[5]}} & \frac{1}{[2]} & -\sqrt{\frac{[6]}{[5][2]}} &  0  \\ \\
\frac{1}{[5]}\sqrt{\frac{[6]}{[3][2]}} & -\sqrt{\frac{[6]}{[5][2]}} & \frac{[6]}{[5][3]} & \sqrt{\frac{1}{[5]}} \\ \\
\sqrt{\frac{[6][2]}{[5][3]}} & 0 & \sqrt{\frac{1}{[5]}} & 0
\end{array}\right)
\ee
}
related with $U_{[5,2,2]}$ by the simple rotation:
\be
\tilde U_{[5,2,2]} = V^{-1} U_{[5,2,2]} V \nn \\
\nn\\
{\footnotesize
V = \left(\begin{array}{cccc}
1 & 0 & 0 & 0 \\ \\
0 & 1 & 0 & 0 \\ \\
0 & 0 & \sin(x) & \cos(x) \\ \\
0 & 0 & -\cos(x) & \sin(x) \\ \\
\end{array}\right)}
\ee
with $\sin(x) = \sqrt{\frac{q^4+q^2+1+q^3+q}{2(q^4+q^2+1)}}, \ \cos(x) =  \sqrt{\frac{q^4+q^2+1-q^3-q}{2(q^4+q^2+1)}} $.

\bigskip

This new mixing matrix can be further decomposed into the product of elementary $2\times 2$ matrices of rotation:
\be
\tilde U_{[5,2,2]} =  r_1r_2r_3r_4r_5r_6,
\ee
where
{\footnotesize
\be
r_1 = \left(\begin{array}{cccc}
1 & 0 & 0 & 0  \\ \\
0 & 1 & 0 & 0  \\ \\
0 & 0 & \cos(x_1) & -\sin(x_1) \\ \\
0 & 0 & \sin(x_1) & \cos(x_1)  \\ \\
\end{array}\right), \
r_2 = \left(\begin{array}{cccc}
1 & 0 & 0 & 0  \\ \\
0 & \cos(x_2) & 0 & \sin(x_2)  \\ \\
0 & 0 & 1 & 0 \\ \\
0 & -\sin(x_2) & 0 & \cos(x_2)  \\ \\
\end{array}\right), \
r_3 = \left(\begin{array}{cccc}
1 & 0 & 0 & 0  \\ \\
0 & \cos(x_3) & -\sin(x_3) & 0  \\ \\
0 & \sin(x_3) & \cos(x_3)  & 0 \\ \\
0 & 0 & 0 & 1 \\ \\
\end{array}\right) \\
r_4 = \left(\begin{array}{cccc}
\cos(x_4) & 0 & 0 & -\sin(x_4)  \\ \\
0 & 1 & 0 & 0  \\ \\
0 & 0 & 1 & 0 \\ \\
\sin(x_4) & 0 & 0 & \cos(x_4)  \\ \\
\end{array}\right), \
r_5 = \left(\begin{array}{cccc}
\cos(x_5) & 0 & \sin(x_5) & 0  \\ \\
0 & 1 & 0 & 0  \\ \\
-\sin(x_5) & 0 &  \cos(x_5) & 0  \\ \\
0 & 0 & 0 & 1  \\ \\
\end{array}\right), \
r_6 = \left(\begin{array}{cccc}
\cos(x_6) & -\sin(x_6) & 0 & 0  \\ \\
\sin(x_6) & \cos(x_6)  & 0 & 0 \\ \\
0 & 0 & 1 & 0 \\ \\
0 & 0 & 0 & 1 \\ \\
\end{array}\right)
\ee
}
and the corresponding angles are
\be
\cos(x_1) = \sqrt{\frac{1}{[5]}} , \hspace{0.5cm}
\cos(x_2) = \frac{[4]}{[2]\sqrt{[5]}}, \hspace{0.5cm}
\cos(x_3) = \frac{1}{[4]}, \hspace{0.5cm}
&\cos(x_4) =  1, \hspace{0.5cm}
\cos(x_5) = 0, \hspace{0.5cm}
\cos(x_6) = \frac{[2]}{[4]} \hspace{0.5cm}\nn \\
\sin(x_1) = \sqrt{\frac{[2][6]}{[3][5]}} , \hspace{0.5cm}
\sin(x_2) = \sqrt{\frac{[6]}{[2][3][5]}}, \hspace{0.5cm}
\sin(x_3) = \frac{\sqrt{[3][5]}}{[4]}, \hspace{0.5cm}
&\sin(x_4) =  0, \hspace{0.5cm}
\sin(x_5) = 1, \hspace{0.5cm}
\sin(x_6) = \frac{\sqrt{[2][6]}}{[4]} \hspace{0.5cm}
\ee

\section{Attaching fingers}

The ingredients that we constructed in the previous sections, that is, the ${\cal R}$-matrices and the 3-strand mixing matrices allow one to immediately evaluate the HOMFLY polynomials of knots/links presented by 3-strand braids via the product of matrices:
\be\label{3strand}
H^{\cal K}_{[2,1]}=\sum_Q \chi^*_Q \cdot \Tr \left({\cal R}_Q^{a_1}\left({\cal U}_Q{\cal R}_Q\widetilde {\cal U}_Q\right)^{b_1}{\cal R}_Q^{a_2}\left({\cal U}_Q{\cal R}_Q\widetilde {\cal U}_Q\right)^{b_2}\ldots\right)
\ee
where $\Tr$ is just the matrix trace, tilde means matrix transposition and the sum goes over all $Q$ lying in $[2,1]^{\otimes 3}$.

However, the variety of knots/links that can be described by a closed 3-strand braid is not that large. It can be considerably enlarged by allowance to insert instead of ${\cal R}$-matrices arbitrary "double-fat fingers" $F_X$, $X\in R\otimes R$. The finger is the building block $B_{X,Y}$, Figure 1 with the top external double-lines closed with each other, i.e. $F_X=B_{0,X}$, the details can be found in \cite{mmmrs}, where the fingers under consideration are called parallel. It is nothing but the plat representation of the two-bridge knot with the two ending arcs cut\footnote{In fact, the class of graphs under consideration is immediately extended in order to include also those with full propagators $B_{Y,X}$, see \cite{MM3str}.
}. The propagator ${B}_{X,Y}$ can be also defined as a 4-strand braid with two parallel and two antiparallel strands.

\begin{figure}
\begin{picture}(300,125)(-200,-30)
\put(0,0){\line(1,0){30}}
\put(0,0){\line(0,1){60}}
\put(0,60){\line(1,0){30}}
\put(30,0){\line(0,1){60}}
\put(2,2){\line(1,0){26}}
\put(2,2){\line(0,1){56}}
\put(2,58){\line(1,0){26}}
\put(28,2){\line(0,1){56}}
\put(5,0){\line(-1,-2){9.5}}
\put(8,0){\line(-1,-2){10}}
\put(25,0){\line(1,-2){9.5}}
\put(22,0){\line(1,-2){10}}
\put(5,60){\line(-1,2){9.5}}
\put(8,60){\line(-1,2){10}}
\put(25,60){\line(1,2){9.5}}
\put(22,60){\line(1,2){10.0}}
\put(-15,-22){\mbox{$X$}}
\put(38,-22){\mbox{$\bar X$}}
\put(-15,80){\mbox{$Y$}}
\put(38,80){\mbox{$\bar Y$}}
\put(7,28){\mbox{$B_{Y\!X}$}}
\end{picture}
\caption{An elementary building block (propagator) for fingers.}
\end{figure}
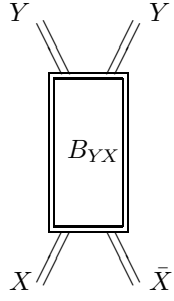

\noindent
In the case of two strand braids, one just inserts $F_X$ instead of ${\cal R}_Q$: the ${\cal R}$-matrix is diagonal and each diagonal entry $R_X$ corresponds to a representation $X\in R\otimes R$. We substitute it with $F_X$ and call the corresponding diagonal matrix ${\cal F}_Q$.
This was done for representation $[2,1]$ in \cite{mmmrs}. In the 3-strand case, one inserts ${\cal F}_Q$ and ${\cal U}_Q{\cal F}_Q\widetilde{\cal U}_Q$. This is what was done in \cite{MM3str} for simpler representations and what we do here for representation $[2,1]$.

The fingers can be chosen arbitrary. As an example, we considered in \cite{MM3str} a 7-parametric family that involved seven possible fingers, four of them being just pretzel fingers:
\be
F^{1}_X(n) =  \frac{(\bar S\bar T^{n}S)_{_{0,X}}}{S_{_{0, X}}}\nn\\
F^{2}_X(n) = \frac{( S T^2 S^\dagger \bar T^{n_{1,6}}S)_{_{0, X}}}{S_{_{0, X}}}\nn\\
F_{X}^{3,\pm}(n) = \frac{(\bar S\bar T^{\pm 2}\bar S \bar T^{n}S)_{_{0, X}}}{S_{_{0, X}}}
\label{7parfam}
\ee
The matrices $T,\bar T$ are in fact the diagonal ${\cal R}$-matrices in the $\hbox{singlet}\in [2,1]^{\otimes 2}\otimes \overline {[2,1]}^{\otimes 2}$ ($T$ corresponds to first two parallel strands and $\bar T$ to first two antiparallel) and $S,\bar S$ the corresponding mixing matrices. They can be found in Appendix B, see \cite{mmmrs} for details.

Let us stress here that though this case of two parallel and two antiparallel strands may look more complicated, since the mixing matrices manifestly depend on $N$ (or $A = q^N$),
in the way we use them, there is no sum over many $Q\in [2,1]\otimes[2,1]\otimes\overline{[2,1]}$
or $Q\in [2,1]\otimes\overline{[2,1]}\otimes[2,1]$, since
only $Q=[2,1]$ contributes.
Hence, one suffices to calculate only two mixing matrices $S$ and $\bar S$ in contrast with $17$
mixing matrices of the previous section.

Thus, following the example in \cite{MM3str}, instead of (\ref{3strand}) we consider a 7-parametric family (see Figure 2)
\be
\chi^*_{[2,1]}H_{[2,1]}^{(n_1,\ldots,n_7|\pm)} \ \ = \nn
\ee
\be\label{7}
=\ \ \sum_{Q\in [2,1]^{\otimes 2}}\chi^*_{Q} \cdot \Tr
\left[{\cal F}_Q^1(n_1){\cal F}_Q^2(n_2)\left({\cal U}_Q{\cal F}_Q^2(n_3)\widetilde {\cal U}_Q\right)
{\cal F}_Q^2(n_4)\left({\cal U}_Q{\cal F}_Q^2(n_5)\widetilde {\cal U}_Q\right){\cal F}_Q^1(n_6)
\left({\cal U}_Q{\cal F}_{Q}^{3,\pm}(n_7)\widetilde {\cal U}_Q\right)\right]
\ee
This family covers almost all the knots from the Rolfsen table, \cite{katlas}. We discussed these HOMFLY polynomials in the next section.

\begin{figure}
\begin{picture}(300,100)(-20,-50)
\put(40,30){\vector(1,0){35}}
\put(40,0){\vector(1,0){35}}
\put(40,-30){\vector(1,0){35}}
\put(75,-30){\vector(1,0){80}}
\qbezier(80,25)(80,35)(85,35)\qbezier(85,35)(90,35)(90,25)
\put(86,35){\vector(1,0){2}}
\qbezier(75,0)(80,0)(80,5)
\put(75,25){\line(1,0){20}}
\put(75,5){\line(0,1){20}}
\put(80,12){\mbox{$n_1$}}
\put(75,5){\line(1,0){20}}
\put(95,5){\line(0,1){20}}
\qbezier(90,5)(90,0)(95,0)
\put(75,30){\vector(1,0){40}}
\put(95,0){\vector(1,0){20}}
\qbezier(115,30)(120,30)(120,25)
\qbezier(115,0)(120,0)(120,5)
\put(115,25){\line(1,0){20}}
\put(115,5){\line(0,1){20}}
\put(120,12){\mbox{$n_2$}}
\put(115,5){\line(1,0){20}}
\put(135,5){\line(0,1){20}}
\qbezier(130,25)(130,30)(135,30)
\qbezier(130,5)(130,0)(135,0)
\put(135,30){\vector(1,0){20}}
\put(135,0){\vector(1,0){20}}
\qbezier(155,-30)(160,-30)(160,-25)
\qbezier(155,0)(160,0)(160,-5)
\put(155,-25){\line(1,0){20}}
\put(155,-5){\line(0,-1){20}}
\put(160,-17){\mbox{$n_3$}}
\put(155,-5){\line(1,0){20}}
\put(175,-5){\line(0,-1){20}}
\qbezier(170,-25)(170,-30)(175,-30)
\qbezier(170,-5)(170,0)(175,0)
\put(155,30){\line(1,0){40}}
\put(175,-30){\vector(1,0){60}}
\put(175,0){\vector(1,0){20}}
\qbezier(195,30)(200,30)(200,25)
\qbezier(195,0)(200,0)(200,5)
\put(195,25){\line(1,0){20}}
\put(195,5){\line(0,1){20}}
\put(200,12){\mbox{$n_4$}}
\put(195,5){\line(1,0){20}}
\put(215,5){\line(0,1){20}}
\qbezier(210,25)(210,30)(215,30)
\qbezier(210,5)(210,0)(215,0)
\put(215,30){\vector(1,0){150}}
\put(215,0){\vector(1,0){20}}
\qbezier(235,-30)(240,-30)(240,-25)
\qbezier(235,0)(240,0)(240,-5)
\put(235,-25){\line(1,0){20}}
\put(235,-5){\line(0,-1){20}}
\put(240,-17){\mbox{$n_5$}}
\put(235,-5){\line(1,0){20}}
\put(255,-5){\line(0,-1){20}}
\qbezier(250,-25)(250,-30)(255,-30)
\qbezier(250,-5)(250,0)(255,0)
\put(255,-30){\vector(1,0){60}}
\put(255,0){\vector(1,0){20}}
%

\qbezier(280,25)(280,35)(285,35)\qbezier(285,35)(290,35)(290,25)
\put(286,35){\vector(1,0){2}}
\qbezier(275,0)(280,0)(280,5)
\put(275,25){\line(1,0){20}}
\put(275,5){\line(0,1){20}}
\put(280,12){\mbox{$n_6$}}
\put(275,5){\line(1,0){20}}
\put(295,5){\line(0,1){20}}
\qbezier(290,5)(290,0)(295,0)
%
\put(295,0){\vector(1,0){20}}
\qbezier(320,-25)(320,-35)(325,-35)\qbezier(325,-35)(330,-35)(330,-25)
\put(324,-35){\vector(-1,0){2}}
\qbezier(315,0)(320,0)(320,-5)
\put(315,-25){\line(1,0){20}}
\put(315,-5){\line(0,-1){20}}
\put(320,-17){\mbox{$n_7$}}
\put(315,-5){\line(1,0){20}}
\put(335,-5){\line(0,-1){20}}
\qbezier(330,-5)(330,0)(335,0)
%
\put(315,-30){\vector(1,0){50}}
\put(335,0){\vector(1,0){30}}
\put(273,30){\vector(1,0){2}}
\end{picture}
\caption{7-parametric family that covers most of the knots in the Rolfsen table.}
\end{figure}
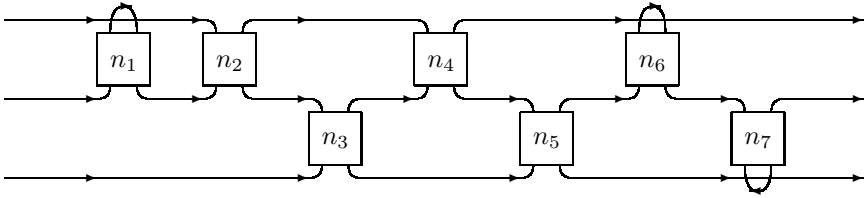

 A few comments are in order. First of all, in all manifest further formulas we use the topological framing, which means that one has renormalize all the ${\cal R}$-matrices with the factor $A^3$ (and with $A^{|R|}$ in the generic representation $R$, cf. (\ref{skein})). Second, we normalize our HOMFLY polynomials $H_R$ dividing by the corresponding HOMFLY polynomial of the unknot, which is nothing but the quantum dimensions $\chi^\ast_R$. Third, in the examples of the knots represented by a 3-braid with fingers which we considered so far, we did not need any additional sign factors even for the knots that unavoidably required introducing non-trivial sign factors in \cite{mmmrs}. The reason is that contributions of the finger components that required non-trivial sign factors to be treated correctly in \cite{mmmrs} (components non-diagonal in multiplicities) do not contribute in the case of fingers of type (\ref{7parfam}).

\section{Polynomials}

As we already mentioned, the 7-parametric family, Figure 2 covers almost the whole Rolfsen table, \cite{katlas} as was demonstrated in \cite[Appendix]{MM3str}. Using (\ref{7}) and the manifest formulas for the mixing matrices, s.3 and Appendix A, one can calculate the colored HOMFLY polynomials in representation $[2,1]$ for these knots. The answers will appear in \cite{ruskatlas}. In fact,
most of data concerning the $[2,1]$-colored HOMFLY polynomials will be appearing in \cite{ruskatlas},
and some has been already published in \cite{AnoMcabling,GJ,mmmrs,nrs}. Here for an illustrative purpose we just give an example of the $[2,1]$-colored HOMFLY polynomial for the enough complicated knot with 10 crossings, $10_{89}$ which is described by at least a 5-strand closed braid (without fingers)\footnote{Evaluating $H_{[21]}$ for 5-strand closed braids looks currently unaffordable.}, and has never been evaluated before:
{\tiny
\be
H_{[21]}^{10_{89}} =
\frac{1}{q^{30}}\cdot\left[
\big(-q^{20}+8q^{18}-27q^{16}+57q^{14}-101q^{12}+169q^{10}-253q^{8}+332q^{6}-399q^{4}+454q^{2}-477+454q^{-2}-399q^{-4}+332q^{-6}
-253q^{-8}
+169q^{-10}-\nn\right.\\-101q^{-12}+57q^{-14}-27q^{-16}+8q^{-18}-q^{-20}\big)+{1\over A^{2}}
\big(q^{26}-7q^{24}+20q^{22}-35q^{20}+55q^{18}-92q^{16}+
140q^{14}-185q^{12}+241q^{10}-315q^{8}+379q^{6}-422q^{4}+\nn\\+462q^{2}-484+462q^{-2}-422q^{-4}+379q^{-6}-315q^{-8}+241q^{-10}-185q^{-12}
+140q^{-14}-92q^{-16}+55q^{-18}-35q^{-20}+20q^{-22}-7q^{-24}+q^{-26}\big)+\nn\\+{1\over A^{4}}
\big(-q^{30}+8q^{28}-32q^{26}+92q^{24}-217q^{22}+
451q^{20}-857q^{18}+1498q^{16}-2421q^{14}+3640q^{12}-5110q^{10}+6714q^{8}-8295q^{6}+9645q^{4}-10537q^{2}+10841-\nn\\-10537q^{-2}+9645q^{-4}
-8295q^{-6}+6714q^{-8}-5110q^{-10}+3640q^{-12}-2421q^{-14}+1498q^{-16}-857q^{-18}+451q^{-20}-217q^{-22}+92q^{-24}-32q^{-26}+8q^{-28}
-\nn\\-q^{-30}\big)+{1\over A^{6}}
\big(4q^{30}-30q^{28}+111q^{26}-298q^{24}+684q^{22}-1410q^{20}+2623q^{18}-4446q^{16}+6974q^{14}-10201q^{12}+
13956q^{10}-17932q^{8}+21741q^{6}-24913q^{4}+\nn\\+27008q^{2}-27736+27008q^{-2}-24913q^{-4}+21741q^{-6}-17932q^{-8}+13956q^{-10}-
10201q^{-12}+6974q^{-14}-4446q^{-16}+2623q^{-18}-1410q^{-20}+684q^{-22}-\nn\\-298q^{-24}+111q^{-26}-30q^{-28}+4q^{-30}\big)+
{1\over A^{8}}
\big(-6q^{30}+40q^{28}-140q^{26}+372q^{24}-850q^{22}+1723q^{20}-3143q^{18}+5251q^{16}-8121q^{14}+11694q^{12}-15804q^{10}+\nn\\+
20148q^{8}-24233q^{6}+27575q^{4}-29822q^{2}+30632-29822q^{-2}+27575q^{-4}-24233q^{-6}+20148q^{-8}-15804q^{-10}+11694q^{-12}-
8121q^{-14}+5251q^{-16}-\nn\\-3143q^{-18}+1723q^{-20}-850q^{-22}+372q^{-24}-140q^{-26}+40q^{-28}-6q^{-30}\big)+{1\over A^{10}}
\big(4q^{30}-
22q^{28}+76q^{26}-208q^{24}+470q^{22}-930q^{20}+1676q^{18}-2766q^{16}+\nn\\+4201q^{14}-5990q^{12}+8048q^{10}-10160q^{8}+
12146q^{6}-13825q^{4}+14904q^{2}-15260+14904q^{-2}-13825q^{-4}+12146q^{-6}-10160q^{-8}+8048q^{-10}-5990q^{-12}+\nn\\+
4201q^{-14}-2766q^{-16}+
1676q^{-18}-930q^{-20}+470q^{-22}-208q^{-24}+76q^{-26}-22q^{-28}+4q^{-30}\big)+{1\over A^{12}}
\big(-q^{30}+4q^{28}-19q^{26}+56q^{24}-
122q^{22}+233q^{20}-\nn\\-396q^{18}+604q^{16}-849q^{14}+1103q^{12}-1323q^{10}+1519q^{8}-1660q^{6}+1734q^{4}-1775q^{2}+1801-1775q^{-2}+
1734q^{-4}-1660q^{-6}+1519q^{-8}-1323q^{-10}+\nn\\+1103q^{-12}-849q^{-14}+604q^{-16}-396q^{-18}+233q^{-20}-122q^{-22}+56q^{-24}-19q^{-26}+
4q^{-28}-q^{-30}\big)+{1\over A^{14}}
\big(3q^{26}-7q^{24}+18q^{22}-30q^{20}+30q^{18}-\nn\\-17q^{16}-13q^{14}+96q^{12}-239q^{10}+396q^{8}-566q^{6}+
742q^{4}-862q^{2}+892-862q^{-2}+742q^{-4}-566q^{-6}+396q^{-8}-239q^{-10}+96q^{-12}-13q^{-14}-17q^{-16}+\nn\\+30q^{-18}-30q^{-20}+18q^{-22}
-7q^{-24}+3q^{-26}\big)+{1\over A^{16}}
\big(-3q^{22}-q^{20}+4q^{18}-12q^{16}+37q^{14}-67q^{12}+85q^{10}-134q^{8}+176q^{6}-170q^{4}+194q^{2}-
230+\nn\\+194q^{-2}-170q^{-4}+176q^{-6}-134q^{-8}+85q^{-10}-67q^{-12}+37q^{-14}-12q^{-16}+4q^{-18}-q^{-20}-3q^{-22}\big)+{1\over A^{18}}
\big(q^{18}
+7q^{16}-5q^{14}+6q^{12}-18q^{10}+15q^{8}-\nn\\-q^{6}+24q^{4}-17q^{2}-4-17q^{-2}+24q^{-4}-q^{-6}+15q^{-8}-18q^{-10}+6q^{-12}-5q^{-14}+
7q^{-16}+q^{-18}\big)+{1\over A^{20}}
\big(-3q^{12}-5q^{10}+2q^{8}+2q^{6}+8q^{4}-10q^{2}-3-\nn\\
\left.-10q^{-2}+8q^{-4}+2q^{-6}+2q^{-8}-5q^{-10}-3q^{-12}\big)+
{1\over A^{22}}\big(3q^{6}+q^{4}+q^{2}-4+q^{-2}+q^{-4}+3q^{-6}\big)-{1\over A^{24}}\right]\nn
\ee
}
Note that a bit more compact is to tabulate not the polynomials themselves, but somewhat more concise data,
polynomials $G_R(A,q)$, an abbreviation made possible by existence of differential expansions
\cite{DGR,evo,arthdiff,AnoMMM21,Kondef}
\be\label{de}
\begin{array}{rl}
H_{[1]} = &1 + G_1\cdot \{Aq\}\{A/q\},\\ \\
H_{[2]} = &1 + [2]G_1\cdot\{Aq^2\}\{A/q\} + G_2\cdot\{Aq^3\}\{Aq^2\}\{A/q\},  \\ \\
H_{[3]} = &1 + [3]G_1\cdot\{Aq^3\}\{A/q\}+ [3]G_2\cdot\{Aq^4\}\{Aq^3\}\{A/q\}+G_3\cdot\{Aq^5\}\{Aq^4\}\{Aq^3\}\{A/q\},\\ \\
H_{[2,1]} = &1+ G_1\cdot\Big(\{Aq^3\}\{A/q^3\}+\{Aq^2\}\{A\}+\{A\}\{A/q^2\}\Big) +  \\
&+\ \{Aq^2\}\{A/q^2\}\cdot\Big([3]\{A\}\cdot G_2(q=1 ) + \{Aq^3\}\{A/q^3\}\cdot G_3(q=1 )
+ \{q\}^2\cdot G_{21}\Big)
\end{array}
\ee
Note that $G_1(A,-q^{-1})=G_1(A,q)$, but $G_2$ and $G_3$ do not obligatory possess this symmetry,
therefore, the decomposition of $H_{[2,1]}$ involves their values at $q=1$.
This choice of $G_{21}$ does not look canonical and can deserve modification.

We remind that for the transposed Young diagram
\be
H_{\tilde R}(A,q) = H_{R}(A,-q^{-1})
\ee
and for the mirror knot
\be
H_R^{\tilde{\cal K}}(A,q) = H_R^{\cal K}(-A^{-1},q)
\ee

If the knot has zero {\it defect} \cite{Kondef}, for example, belongs to the twist-knots family,
then $G_1$ is independent of $q$, while
$G_2$ and $G_3$ are further factorized: they are divisible by $\{A\}$ and $\{A\}\{Aq\}$ respectively,
then $G_{21}$ is also divisible by $\{A\}$ (this follows from (\ref{de}), since at $A=1$ ones gets the Alexander polynomial $Al_R(q)$, which celebrates the property $Al_R(q)=Al_{[1]}(q^{|R|})$ for any hook diagram $R$, \cite{IMMMfe}).

As an illustration of this kind decomposition, we write it for knot $8_1$ which is the twist knot with twist=-3 and which is described by at least a 5-strand closed braid:
\be
G_1^{8_1} = \frac{A^4+A^2+1}{A^4},
\nn \\ \nn \\
G_2^{8_1} = \{A\}\frac{A^8q^8+A^6q^8+A^6q^6+A^4q^8+A^4q^6+A^4q^4+A^2q^4+A^2q^2+1}{A^8q^8},
\nn \\ \nn \\
G_3^{8_1} = \{A\}\{Aq\}A^{-12}q^{-24}\Big(A^{12}q^{24}+A^{10}q^{24}+A^{10}q^{22}+A^{10}q^{20}+A^8q^{24}
+A^8q^{22}
+2A^8q^{20}+A^8q^{18}+A^8q^{16}+A^6q^{20}+\nn\\
+2A^6q^{18}+2A^6q^{16}+A^6q^{14}+A^6q^{12}+A^4q^{16}+A^4q^{14}
+2A^4q^{12}+A^4q^{10}+A^4q^8+A^2q^8+A^2q^6+A^2q^4+1\Big),
\nn \\ \nn \\
G_{21}^{8_1} = \{A\}A^{-13}q^{-12}\Big(-7A^6q^8+A^4q^8-4A^8q^6-4A^6q^6+2A^4q^4-A^6q^2-A^{10}q^{20}
-2A^8q^{20}-A^{12}q^{18}-2A^{10}q^{18}-\nn\\
-4A^8q^{18}-4A^6q^{18}-2A^{10}q^{16}-4A^8q^{16}-5A^8q^{14} -8A^6q^{14}
-4A^8q^{12}-9A^6q^{12}-5A^8q^{10}-8A^6q^{10}-2A^{10}q^8-\nn\\
-4A^8q^8-2A^8q^4-3A^6q^4-A^8q^2-2A^{10}q^6
-A^{10}q^4-6A^4q^{10}-A^4q^6+A^2q^4-6A^4q^{14}-3A^2q^{14}-2A^4q^{12}-\nn\\
-2A^2q^{12}-3A^2q^{10}+A^{14}q^{14}
+A^{14}q^{12}+2A^{12}q^{12}+A^{14}q^{10}-A^{12}q^6+A^4-q^{14}-q^{12}-q^{10}-A^6+A^2q^{20}+\nn \\
+A^4q^{24}-A^6q^{24}-A^6q^{22}-3A^6q^{20}+2A^4q^{20}-A^4q^{18}+A^4q^{16}-A^8q^{22}-7A^6q^{16}\Big)
\nn
\ee

\section*{Acknowledgements}

This work is a part of a large project that has been started together by the present authors along with Ramadevi Pichai and Vivek  Kumar Singh and includes also Petr Dunin-Barkowski and Alexander Popolitov. We are grateful to them for fruitful discussions and continuing collaboration.

Our work is partly supported by grants NSh-1500.2014.2, by RFBR grants 13-02-00457 (A.Mir. \& A.S.), 13-02-00478 (A.Mor.), 14-01-00547 (An.Mor.), by the joint grants 15-52-50034-YaF, 15-51-52031-NSC-a, by 14-01-92691-Ind-a, by grants 14-01-31395-Mol-a (An.M.),
14-02-31446-Mol-a (A.S.) and 15-31-20832-Mol-a-ved (A.M. \& A.S.). Also we are partly supported by
the Brazilian National Counsel of Scientific and Technological Development (A.Mor.)
and by the Quantum Topology Lab of Chelyabinsk State University (Russian Federation government grant 14.Z50.31.0020) (An.M. \& A.S.).

\newpage

\section*{Appendix A}

Here we list the remaining ${\cal R}$-matrices and mixing matrices that we calculated directly from highest weight vectors, see s.2:

{\footnotesize
\be
{\cal R}_{[5,3,1]} = \left(\begin{array}{cccccc}
q^{-\varkappa_{[4,2]}} &&&&&\\ \\
& q^{-\varkappa_{[4,2]}} &&&& \\ \\
&& -q^{-\varkappa_{[4,1,1]}} &&& \\ \\
&&& -q^{-\varkappa_{[3,3]}} &&\\ \\
&&&& -q^{-\varkappa_{[3,2,1]}} & \\ \\ &&&&& q^{-\varkappa_{[3,2,1]}}
\end{array}\right)
= \left(\begin{array}{cccccc}
\frac{1}{q^5} &&&&& \\ \\ & \frac{1}{q^5} &&&& \\ \\
&& -\frac{1}{q^3} &&& \\ \\  &&& -\frac{1}{q^3} && \\ \\
&&&&-1& \\ \\ &&&&&1
\end{array}\right)
\ee
\be
{\cal U}_{[5,3,1]} = \left(\begin{array}{cccccc}
-\frac{[9]+[5]+2[3]-1}{[5][2]\a_1} & -\frac{\a_0}{\a_1}\sqrt{\frac{[7][3]}{[5]}} & \frac{\a_0^2}{\sqrt{ [5][3]\a_1 }} &  \frac{\sqrt{ \a_1 }}{[2]\sqrt{ [5][3] }} & X_{15} & X_{16} \\ \\
-\frac{\a_0}{\a_1}\sqrt{\frac{[7][3]}{[5]}} & \frac{[3]^2}{[2]\a_1} & \frac{\sqrt{[7]}}{[2]\sqrt{\a_1}} & 0 & X_{25} & X_{26} \\ \\
\frac{\a_0^2}{\sqrt{ [5][3]\a_1 }} & \frac{\sqrt{[7]}}{[2]\sqrt{\a_1}} & \frac{1}{[2][3]} & \frac{\a_0}{[3]} & X_{35} & X_{36} \\ \\
\frac{\sqrt{ \a_1 }}{[2]\sqrt{ [5][3] }} & 0 & \frac{\a_0}{[3]} & \frac{1}{[2][3]} & X_{45} & X_{46} \\ \\
X_{15} & X_{25} & X_{35} & X_{45} & X_{55} & X_{56} \\ \\
X_{16} & X_{26} & X_{36} & X_{46} & X_{56} & X_{66}
\end{array}\right)
\ee
with $\a_1=[7]+[5]+1, \ \a_0=[3]-1$.

\be
X_{15} = -1/2\,{\frac {\sqrt {{q}^{4}+1}\sqrt {2} \left( {q}^{6}-{q}^{5}+{q}^{4
}-{q}^{3}+{q}^{2}-q+1 \right) \sqrt {{q}^{2}-q+1}\sqrt {{q}^{4}+{q}^{2
}+1+{q}^{3}+q}}{\sqrt {{q}^{2}+q+1} \left( {q}^{4}+{q}^{2}+1-{q}^{3}-q
 \right) {q}^{3}\sqrt {\alpha_{{1}}}}} \nn \\
X_{16} = 1/2\,{\frac {\sqrt {{q}^{4}+1}\sqrt {2} \left( {q}^{6}+{q}^{5}+{q}^{4}
+{q}^{3}+{q}^{2}+q+1 \right) \sqrt {{q}^{2}+q+1}\sqrt {{q}^{4}+{q}^{2}
+1-{q}^{3}-q}}{\sqrt {{q}^{2}-q+1} \left( {q}^{4}+{q}^{2}+1+{q}^{3}+q
 \right) {q}^{3}\sqrt {\alpha_{{1}}}}} \nn \\
X_{25} = 1/2\,{\frac {\sqrt {{q}^{4}+1}\sqrt {{q}^{6}+{q}^{5}+{q}^{4}+{q}^{3}+{
q}^{2}+q+1}\sqrt {{q}^{6}-{q}^{5}+{q}^{4}-{q}^{3}+{q}^{2}-q+1}\sqrt {2
}}{\sqrt {{q}^{4}+{q}^{2}+1-{q}^{3}-q}{q}^{3}\sqrt {\alpha_{{1}}}}}
\nn \\
X_{26} = 1/2\,{\frac {\sqrt {{q}^{4}+1}\sqrt {{q}^{6}+{q}^{5}+{q}^{4}+{q}^{3}+{
q}^{2}+q+1}\sqrt {{q}^{6}-{q}^{5}+{q}^{4}-{q}^{3}+{q}^{2}-q+1}\sqrt {2
}}{\sqrt {{q}^{4}+{q}^{2}+1+{q}^{3}+q}{q}^{3}\sqrt {\alpha_{{1}}}}}
 \nn \\
X_{35} = -1/2\,{\frac {q\sqrt {2}\sqrt {{q}^{4}+1}}{\sqrt {{q}^{4}+{q}^{2}+1-{q
}^{3}-q} \left( {q}^{2}+q+1 \right) }}, \ X_{36} = 1/2\,{\frac {q\sqrt {2}\sqrt {{q}^{4}+1}}{\sqrt {{q}^{4}+{q}^{2}+1+{q}
^{3}+q} \left( {q}^{2}-q+1 \right) }} \nn \\
X_{45} = 1/2\,{\frac {q\sqrt {2}\sqrt {{q}^{4}+1}}{\sqrt {{q}^{4}+{q}^{2}+1-{q}
^{3}-q} \left( {q}^{2}+q+1 \right) }}, \ X_{46} = 1/2\,{\frac {q\sqrt {2}\sqrt {{q}^{4}+1}}{\sqrt {{q}^{4}+{q}^{2}+1+{q}^{3}+q} \left( {q}^{2}-q+1 \right) }} \nn
\ee
\be
X_{55} = -{\frac {{q}^{3}}{ \left( {q}^{2}+q+1 \right)  \left( {q}^{4}+{q}^{2}+
1-{q}^{3}-q \right) }}, \ X_{56} = 0 \nn \\
X_{65} = 0, \ X_{66} = -{\frac {{q}^{3}}{ \left( {q}^{2}-q+1 \right)  \left( {q}^{4}+{q}^{2}+1+{q}^{3}+q \right) }} \nn
\ee

\be
{\cal R}_{[5,2,1,1]} = \left(\begin{array}{cccccc}
q^{-\varkappa_{[4,2]}} &&&&&\\ \\
& -q^{-\varkappa_{[4,1,1]}} &&&& \\ \\
&& -q^{-\varkappa_{[4,1,1]}} &&& \\ \\
&&&  -q^{-\varkappa_{[3,2,1]}} &&\\ \\
&&&& q^{-\varkappa_{[3,2,1]}} & \\ \\ &&&&& q^{-\varkappa_{[3,1,1,1]}}
\end{array}\right)
= \left(\begin{array}{cccccc}
\frac{1}{q^5} &&&&& \\ \\ & -\frac{1}{q^3} &&&& \\ \\
&& -\frac{1}{q^3} &&& \\ \\  &&& -1 && \\ \\
&&&&1& \\ \\ &&&&&q^3
\end{array}\right)
\ee
\be
{\cal U}_{[5,2,1,1]} = \left(\begin{array}{cccccc}
\frac{[3]}{[5][2]} & \frac{ 1 }{ [2] } \sqrt{ \frac{[5]-[3]}{[5]\a_0} } & -\sqrt{ \frac{[2]}{[6]\a_0} } & Y_{14} & Y_{15} & \left([3]-1\right)\sqrt{ \frac{([5]-[3])[2]}{[6][5]} } \\ \\
\frac{ 1 }{ [2] } \sqrt{ \frac{[5]-[3]}{[5]\a_0} } & \frac{1}{[2]} & 0 & Y_{24} & Y_{25} & \frac{1}{[3]} \sqrt{ \frac{[6]}{[2]\a_0} } \\ \\
-\sqrt{ \frac{[2]}{[6]\a_0} } & 0 & -\frac{1}{[6]} & Y_{34} & Y_{35} & \frac{1}{[6]} \sqrt{ \frac{[5]([5]-[3])}{[3]-1} } \\ \\
Y_{14} & Y_{24} & Y_{34} & Y_{44} & Y_{45} & Y_{46} \\ \\
Y_{15} & Y_{25} & Y_{35} & Y_{45} & Y_{55} & Y_{56} \\ \\
\left([3]-1\right)\sqrt{ \frac{([5]-[3])[2]}{[6][5]} } & \frac{1}{[3]} \sqrt{ \frac{[6]}{[2]\a_0} } & \frac{1}{[6]} \sqrt{ \frac{[5]([5]-[3])}{[3]-1} } & Y_{64} & Y_{65} &  \frac{1}{[6]}
\end{array}\right)
\ee

\be
Y_{14} = 1/2\,{\frac {\sqrt {{q}^{8}+1}\sqrt {{q}^{4}+1}\sqrt {2}q}{\sqrt {{q}^
{4}+{q}^{2}+1+{q}^{3}+q} \left( {q}^{4}+{q}^{2}+1-{q}^{3}-q \right)
\sqrt {{q}^{2}+q+1}\sqrt {{q}^{2}-q+1}}}
\nn \\
Y_{15} = 1/2\,{\frac {\sqrt {{q}^{8}+1}\sqrt {{q}^{4}+1}\sqrt {2}q}{ \left( {q}
^{4}+{q}^{2}+1+{q}^{3}+q \right) \sqrt {{q}^{4}+{q}^{2}+1-{q}^{3}-q}
\sqrt {{q}^{2}+q+1}\sqrt {{q}^{2}-q+1}}}
\nn \\
Y_{24} = 1/2\,{\frac { \left( q-1 \right) ^{2}\sqrt {{q}^{2}+q+1}\sqrt {2}}{
\sqrt {{q}^{2}-q+1}\sqrt {{q}^{4}+{q}^{2}+1-{q}^{3}-q}}}
, \ Y_{25} = 1/2\,{\frac { \left( 1+q \right) ^{2}\sqrt {{q}^{2}-q+1}\sqrt {2}}{
\sqrt {{q}^{2}+q+1}\sqrt {{q}^{4}+{q}^{2}+1+{q}^{3}+q}}}
 \nn \\
Y_{34} = -1/2\,{\frac {\sqrt {{q}^{8}+1}\sqrt {2}\sqrt {{q}^{4}+{q}^{2}+1+{q}^{
3}+q}}{ \left( {q}^{2}-q+1 \right)  \left( {q}^{2}+q+1 \right) \sqrt {
{q}^{4}-{q}^{2}+1}}}
, \ Y_{35} = 1/2\,{\frac {\sqrt {{q}^{8}+1}\sqrt {2}\sqrt {{q}^{4}+{q}^{2}+1-{q}^{3
}-q}}{ \left( {q}^{2}-q+1 \right)  \left( {q}^{2}+q+1 \right) \sqrt {{
q}^{4}-{q}^{2}+1}}}
\nn \\
Y_{44} = -1/2\,{\frac {q \left( {q}^{6}-2\,{q}^{5}+3\,{q}^{4}-2\,{q}^{3}+3\,{q}
^{2}-2\,q+1 \right) }{ \left( {q}^{2}-q+1 \right)  \left( {q}^{2}+q+1
 \right)  \left( {q}^{4}+{q}^{2}+1-{q}^{3}-q \right) }}
\nn \\
Y_{45} = -1/2\,{\frac { \left( {q}^{2}+1 \right)  \left( {q}^{4}+1 \right) q}{
 \left( {q}^{2}-q+1 \right)  \left( {q}^{2}+q+1 \right) \sqrt {{q}^{4}
+{q}^{2}+1+{q}^{3}+q}\sqrt {{q}^{4}+{q}^{2}+1-{q}^{3}-q}}}
\nn \\
Y_{55} = -1/2\,{\frac {q \left( {q}^{6}+2\,{q}^{5}+3\,{q}^{4}+2\,{q}^{3}+3\,{q}
^{2}+2\,q+1 \right) }{ \left( {q}^{4}+{q}^{2}+1+{q}^{3}+q \right)
 \left( {q}^{2}-q+1 \right)  \left( {q}^{2}+q+1 \right) }}
\nn \\
Y_{64} = -1/2\,{\frac {\sqrt {2}\sqrt {{q}^{4}+1}{q}^{3}}{ \left( {q}^{2}-q+1
 \right)  \left( {q}^{2}+q+1 \right) \sqrt {{q}^{4}+{q}^{2}+1-{q}^{3}-
q}\sqrt {{q}^{4}-{q}^{2}+1}}}
\nn \\
Y_{65} = 1/2\,{\frac {\sqrt {2}\sqrt {{q}^{4}+1}{q}^{3}}{ \left( {q}^{2}-q+1
 \right)  \left( {q}^{2}+q+1 \right) \sqrt {{q}^{4}+{q}^{2}+1+{q}^{3}+
q}\sqrt {{q}^{4}-{q}^{2}+1}}}
\nn
\ee

\be
{\cal R}_{[4,3,2]} = \left(\begin{array}{cccccccc}
q^{-\varkappa_{[4,2]}} &&&&&&&\\ \\
& -q^{-\varkappa_{[4,1,1]}} &&&&&& \\ \\
&& -q^{-\varkappa_{[3,3]}} &&&&& \\ \\
&&&  -q^{-\varkappa_{[3,2,1]}} &&&&\\ \\
&&&& -q^{-\varkappa_{[3,2,1]}} &&& \\ \\
&&&&&  q^{-\varkappa_{[3,2,1]}} &&\\ \\
&&&&&& q^{-\varkappa_{[3,2,1]}} & \\ \\
&&&&&&& q^{-\varkappa_{[2,2,2]}}
\end{array}\right) = \nn \\
= \left(\begin{array}{cccccccc}
\frac{1}{q^5} &&&&&&& \\ \\ & -\frac{1}{q^3} &&&&&& \\ \\
&& -\frac{1}{q^3} &&&&& \\ \\  &&& -1 &&&& \\ \\ &&&&-1 &&& \\ \\
&&&&&1&& \\ \\ &&&&&&1& \\ &&&&&&&q^3
\end{array}\right)
\ee
\be
{\cal U}_{[4,3,2]} = \left(\begin{array}{cccccccc}
\frac{[5]+2}{[5][2]^2\a_0} & \frac{1}{[2]^2}\sqrt{\frac{[6]}{[5][2]\a_0}} & \frac{1}{[2]^2}\sqrt{\frac{[6]}{[5][2]\a_0}} & Z_{14} & 0 & 0 & Z_{17} & \frac{[3]}{[2]^2\a_0} \sqrt{\frac{[6]}{[2]}} \\ \\
\frac{1}{[2]^2}\sqrt{\frac{[6]}{[5][2]\a_0}} & \frac{1}{[2]^2} & \frac{1}{[2]^2} & Z_{24} & \frac{1}{2} & \frac{1}{2} & Z_{27} & -\frac{1}{[2]^2}\sqrt{\frac{[5]}{\a_0}}  \\ \\
\frac{1}{[2]^2}\sqrt{\frac{[6]}{[5][2]\a_0}} & \frac{1}{[2]^2} & \frac{1}{[2]^2} & Z_{34} &-\frac{1}{2} & -\frac{1}{2} & Z_{37} & -\frac{1}{[2]^2}\sqrt{\frac{[5]}{\a_0}}  \\ \\
Z_{14} & Z_{24} & Z_{34} & Z_{44} & 0 & 0 & Z_{74} &  Z_{84} \\ \\
0 & \frac{1}{2} & -\frac{1}{2} & 0 & \frac{1}{2} & -\frac{1}{2} & 0 & 0  \\ \\
0 & \frac{1}{2} & -\frac{1}{2} & 0 & -\frac{1}{2} & \frac{1}{2} & 0 & 0  \\ \\
Z_{17} & Z_{27} & Z_{37} & Z_{47} & 0 & 0 & Z_{77} & Z_{87}  \\ \\
\frac{[3]}{[2]^2\a_0} \sqrt{\frac{[6]}{[2]}} & -\frac{1}{[2]^2}\sqrt{\frac{[5]}{\a_0}} & -\frac{1}{[2]^2}\sqrt{\frac{[5]}{\a_0}} & Z_{84} & 0 & 0 & Z_{87} &  \frac{1}{[2]^2\a_0}
\end{array}\right)
\ee

\be
\begin{array}{ll}
Z_{14}= {\frac {q\sqrt {{q}^{2}+q+1}\sqrt {{q}^{4}-{q}^{2}+1} \left( {q}^{2}-q
+1 \right) ^{3/2}}{\sqrt {{q}^{4}+1} \left( {q}^{2}+1 \right)  \left(
{q}^{4}+{q}^{2}+1-{q}^{3}-q \right) }}
 & Z_{17}= -{\frac {q\sqrt {{q}^{2}-q+1}\sqrt {{q}^{4}-{q}^{2}+1} \left( {q}^{2}+
q+1 \right) ^{3/2}}{\sqrt {{q}^{4}+1} \left( {q}^{2}+1 \right)
 \left( {q}^{4}+{q}^{2}+1+{q}^{3}+q \right) }}
 \nn \\ \nn \\
Z_{24}= 1/2\,{\frac {\sqrt {{q}^{4}+{q}^{2}+1+{q}^{3}+q} \left( q-1 \right) ^{
2}}{\sqrt {{q}^{4}+{q}^{2}+1-{q}^{3}-q} \left( {q}^{2}+1 \right) }}
 & Z_{27}= -1/2\,{\frac {\sqrt {{q}^{4}+{q}^{2}+1-{q}^{3}-q} \left( 1+q \right) ^
{2}}{\sqrt {{q}^{4}+{q}^{2}+1+{q}^{3}+q} \left( {q}^{2}+1 \right) }}
 \nn \\ \nn \\
Z_{34}= 1/2\,{\frac {\sqrt {{q}^{4}+{q}^{2}+1+{q}^{3}+q} \left( q-1 \right) ^{
2}}{\sqrt {{q}^{4}+{q}^{2}+1-{q}^{3}-q} \left( {q}^{2}+1 \right) }}
 & Z_{37}= -1/2\,{\frac {\sqrt {{q}^{4}+{q}^{2}+1-{q}^{3}-q} \left( 1+q \right) ^
{2}}{\sqrt {{q}^{4}+{q}^{2}+1+{q}^{3}+q} \left( {q}^{2}+1 \right) }}
 \nn \\ \nn \\
Z_{44}= 1/2\,{\frac {{q}^{4}+{q}^{3}-3\,{q}^{2}+q+1}{{q}^{4}+{q}^{2}+1-{q}^{3}
-q}} &
 Z_{47}= Z_{74}=\frac{1}{2} \nn\\ \nn\\
Z_{54}= Z_{57}= Z_{64}= Z_{67}= 0 & Z_{77}= 1/2\,{\frac {{q}^{4}-{q}^{3}-3\,{q}^{2}-q+1}{{q}^{4}+{q}^{2}+1+{q}^{3}
+q}}
 \nn \\ \nn \\
Z_{84}= -{\frac {{q}^{2}}{\sqrt {{q}^{4}+1} \left( {q}^{2}+1 \right) }} & Z_{87}= -{\frac {{q}^{2}}{\sqrt {{q}^{4}+1} \left( {q}^{2}+1 \right) }} \nn
\end{array}
\ee

\be
{\cal R}_{[4,3,1,1]} = \left(\begin{array}{ccccccccc}
q^{-\varkappa_{[4,2]}} &&&&&&&&\\ \\
& -q^{-\varkappa_{[4,1,1]}} &&&&&&& \\ \\
&& -q^{-\varkappa_{[3,3]}} &&&&&& \\ \\
&&&  -q^{-\varkappa_{[3,2,1]}} &&&&&\\ \\
&&&&  -q^{-\varkappa_{[3,2,1]}} &&&&\\ \\
&&&&& q^{-\varkappa_{[3,2,1]}} &&& \\ \\
&&&&&& q^{-\varkappa_{[3,2,1]}} && \\ \\
&&&&&&& q^{-\varkappa_{[3,1,1,1]}} & \\ \\
&&&&&&&& -q^{-\varkappa_{[2,2,1,1]}}  \\ \\
\end{array}\right) = \nn
\ee
\be
= \left(\begin{array}{ccccccccc}
\frac{1}{q^5} &&&&&&&& \\ \\ & -\frac{1}{q^3} &&&&&&& \\ \\ && -\frac{1}{q^3} &&&&&& \\ \\
&&& -1 &&&&& \\ \\  &&&& -1 &&&& \\ \\
&&&&&1&&& \\ \\ &&&&&&1 && \\ \\ &&&&&&&q^3 & \\ \\&&&&&&&& -q^5
\end{array}\right)
\ee
\be
{\cal U}_{[4,3,1,1]} = \left(\begin{array}{ccccccccc}
\frac{[3]}{[5][2]^2} & \frac{1}{[2]}\sqrt{\frac{\a_0}{[6][2]}} & \frac{[3]}{[2]^2}\sqrt{\frac{1}{[5]\a_0}}
& W_{1,4} & W_{1,5} & W_{1,6} & W_{1,7} & \frac{1}{[2]}\sqrt{\frac{[7]}{[6][2]}} & \frac{[3]}{[2]^2}\sqrt{\frac{[7]}{[5]\a_0}} \\ \\
\frac{1}{[2]}\sqrt{\frac{\a_0}{[6][2]}} & \frac{\a_2}{[6][2]} & \frac{1}{[2]}\sqrt{\frac{[5]}{[6][2]}}
& W_{2,4} & W_{2,5} & W_{2,6} &  W_{2,7}& -\frac{1}{[6][2]}\sqrt{[7]\a_0} & -\frac{[3]}{[2]}\sqrt{\frac{[7]}{[6][5][2]}} \\ \\
\frac{[3]}{[2]^2}\sqrt{\frac{1}{[5]\a_0}} & \frac{1}{[2]}\sqrt{\frac{[5]}{[6][2]}} & -\frac{1}{[2]^2\a_0}
& W_{3,4} & W_{3,5} & W_{3,6} & W_{3,7} & \frac{1}{[2]}\sqrt{\frac{[7][5]}{[6][2]\a_0}} & -\frac{1}{[2]^2\a_0}\sqrt{[7]} \\ \\
W_{1,4} & W_{2,4} & W_{3,4} & W_{4,4} & W_{4,5} & W_{4,6} & W_{4,7} & W_{8,4} & W_{9,4} \\ \\
W_{1,5} & W_{2,5} & W_{3,5} &  W_{5,4}& W_{5,5} & W_{5,6} & W_{5,7} & W_{8,5} & W_{9,5} \\ \\
W_{1,6} & W_{2,6} & W_{3,6} & W_{6,4} & W_{6,5} & W_{6,6} & W_{6,7} & W_{8,6} & W_{9,6} \\ \\
W_{1,7} & W_{2,7} & W_{3,7} & W_{7,4} & W_{7,5} & W_{7,6} & W_{7,7} & W_{8,7} & W_{9,7} \\ \\
\frac{1}{[2]}\sqrt{\frac{[7]}{[6][2]}} & -\frac{1}{[6][2]}\sqrt{[7]\a_0}
& \frac{1}{[2]}\sqrt{\frac{[7][5]}{[6][2]\a_0}}
& W_{8,4} & W_{8,5} & W_{8,6} & W_{8,7} & \frac{1}{[6][2]} & \frac{[3]}{[2]^2}\sqrt{\frac{[2]}{[6][5]\a_0}} \\ \\
\frac{[3]}{[2]^2}\sqrt{\frac{[7]}{[5]\a_0}} & -\frac{[3]}{[2]}\sqrt{\frac{[7]}{[6][5][2]}}
& -\frac{1}{[2]^2\a_0}\sqrt{[7]}
& W_{9,4} & W_{9,5} & W_{9,6} & W_{9,7} & \frac{[3]}{[2]^2}\sqrt{\frac{[2]}{[6][5]\a_0}} &  -\frac{[3]}{[5][2]^2\a_0}
\end{array}\right)
\ee
$\a_2=[5]-[3]$

\be
W_{{1,4}}=1/2\,{\frac {\sqrt {2}{q}^{2}\sqrt {{q}^{6}+{q}^{5}+{q}^{4}
+{q}^{3}+{q}^{2}+q+1}\sqrt {{q}^{6}-{q}^{5}+{q}^{4}-{q}^{3}+{q}^{2}-q+
1}}{\sqrt {{q}^{4}+{q}^{2}+1+{q}^{3}+q}\sqrt {{q}^{2}-q+1} \left( {q}^
{2}+1 \right) \sqrt {{q}^{2}+q+1} \left( {q}^{4}+{q}^{2}+1-{q}^{3}-q
 \right) }} \nn \\ W_{{2,4}}=1/2\,{\frac {\sqrt {{q}^{6}+{q}^{5}+{q}^{4}+{q}^
{3}+{q}^{2}+q+1}\sqrt {{q}^{6}-{q}^{5}+{q}^{4}-{q}^{3}+{q}^{2}-q+1}q
\sqrt {{q}^{4}+1}\sqrt {2}}{\sqrt {{q}^{4}+{q}^{2}+1+{q}^{3}+q}
 \left( {q}^{2}+1 \right)  \left( {q}^{2}-q+1 \right)  \left( {q}^{2}+
q+1 \right) \sqrt {{q}^{4}-{q}^{2}+1}}} \nn \\ W_{{3,4}}=1/2\,{\frac {\sqrt {
2}q\sqrt {{q}^{6}-{q}^{5}+{q}^{4}-{q}^{3}+{q}^{2}-q+1}\sqrt {{q}^{6}+{
q}^{5}+{q}^{4}+{q}^{3}+{q}^{2}+q+1}}{\sqrt {{q}^{4}+{q}^{2}+1-{q}^{3}-
q}\sqrt {{q}^{2}+q+1}\sqrt {{q}^{2}-q+1} \left( {q}^{2}+1 \right)
\sqrt {{q}^{4}+1}}} \nn \\ W_{{4,4}}=1/2\,{\frac {{q}^{12}-{q}^{11}+{q}^{10}-
{q}^{9}+{q}^{8}-2\,{q}^{7}-2\,{q}^{5}+{q}^{4}-{q}^{3}+{q}^{2}-q+1}{
 \left( {q}^{2}-q+1 \right)  \left( {q}^{2}+q+1 \right)  \left( {q}^{4
}+{q}^{2}+1+{q}^{3}+q \right)  \left( {q}^{4}+{q}^{2}+1-{q}^{3}-q
 \right) }} \nn \\ W_{{5,4}}=-1/2\,{\frac { \left( {q}^{4}+{q}^{3}-{q}^{2}+q+
1 \right) \sqrt {{q}^{6}+{q}^{5}+{q}^{4}+{q}^{3}+{q}^{2}+q+1}\sqrt {{q
}^{6}-{q}^{5}+{q}^{4}-{q}^{3}+{q}^{2}-q+1}}{ \left( {q}^{4}+{q}^{2}+1+
{q}^{3}+q \right) \sqrt {{q}^{2}-q+1}\sqrt {{q}^{2}+q+1} \left( {q}^{4
}+{q}^{2}+1-{q}^{3}-q \right) }} \nn \\ W_{{6,4}}=1/2\,{\frac { \left( {q}^{
10}-2\,{q}^{9}+2\,{q}^{8}-3\,{q}^{7}+2\,{q}^{6}-3\,{q}^{5}+2\,{q}^{4}-
3\,{q}^{3}+2\,{q}^{2}-2\,q+1 \right) \sqrt {{q}^{2}+q+1}}{\sqrt {{q}^{
4}+{q}^{2}+1+{q}^{3}+q}\sqrt {{q}^{4}+{q}^{2}+1-{q}^{3}-q}\sqrt {{q}^{
2}-q+1}\sqrt {{q}^{12}+2\,{q}^{10}+4\,{q}^{8}+5\,{q}^{6}+4\,{q}^{4}+2
\,{q}^{2}+1}}} \nn \\ W_{{7,4}}=-1/2\,{\frac { \left( {q}^{2}+1 \right)
 \left( {q}^{6}+{q}^{5}+{q}^{4}+{q}^{3}+{q}^{2}+q+1 \right) ^{3/2}
\sqrt {{q}^{6}-{q}^{5}+{q}^{4}-{q}^{3}+{q}^{2}-q+1}}{ \left( {q}^{2}-q
+1 \right)  \left( {q}^{2}+q+1 \right) \sqrt {{q}^{4}+{q}^{2}+1+{q}^{3
}+q}\sqrt {{q}^{4}+{q}^{2}+1-{q}^{3}-q}\sqrt {{q}^{12}+2\,{q}^{10}+4\,
{q}^{8}+5\,{q}^{6}+4\,{q}^{4}+2\,{q}^{2}+1}}} \nn \\ W_{{8,4}}=-1/2\,{\frac {
\sqrt {2}{q}^{5}}{\sqrt {{q}^{4}+{q}^{2}+1+{q}^{3}+q} \left( {q}^{2}+1
 \right)  \left( {q}^{2}-q+1 \right)  \left( {q}^{2}+q+1 \right)
\sqrt {{q}^{4}-{q}^{2}+1}}} \nn \\ W_{{9,4}}=-1/2\,{\frac { \left( {q}^{2}+q+
1 \right) ^{3/2}\sqrt {2}{q}^{3}\sqrt {{q}^{2}-q+1}}{ \left( {q}^{4}+{
q}^{2}+1+{q}^{3}+q \right)  \left( {q}^{2}+1 \right) \sqrt {{q}^{4}+{q
}^{2}+1-{q}^{3}-q}\sqrt {{q}^{4}+1}}} \nn
\ee
\be
W_{{1,5}}=1/2\,{\frac { \left( {q}^{6}-{q}^{5}+{q}^{4}-{q}^{3}+{q}^{2
}-q+1 \right) q\sqrt {2}}{ \left( {q}^{4}+{q}^{2}+1-{q}^{3}-q \right)
\sqrt {{q}^{4}+{q}^{2}+1+{q}^{3}+q} \left( {q}^{2}+1 \right) }} \nn \\ W_{{2,
5}}=1/2\,{\frac { \left( {q}^{6}+{q}^{3}+1 \right) \sqrt {{q}^{4}+1}
\sqrt {2}}{\sqrt {{q}^{4}+{q}^{2}+1+{q}^{3}+q}\sqrt {{q}^{2}-q+1}
 \left( {q}^{2}+1 \right) \sqrt {{q}^{2}+q+1}\sqrt {{q}^{4}-{q}^{2}+1}
}} \nn \\ W_{{3,5}}=-1/2\,{\frac { \left( {q}^{2}-q+1 \right) {q}^{2}\sqrt {2
}}{\sqrt {{q}^{4}+{q}^{2}+1-{q}^{3}-q} \left( {q}^{2}+1 \right) \sqrt
{{q}^{4}+1}}} \nn \\ W_{{4,5}}=-1/2\,{\frac { \left( {q}^{4}+{q}^{3}-{q}^{2}+
q+1 \right) \sqrt {{q}^{6}+{q}^{5}+{q}^{4}+{q}^{3}+{q}^{2}+q+1}\sqrt {
{q}^{6}-{q}^{5}+{q}^{4}-{q}^{3}+{q}^{2}-q+1}}{ \left( {q}^{4}+{q}^{2}+
1+{q}^{3}+q \right) \sqrt {{q}^{2}-q+1}\sqrt {{q}^{2}+q+1} \left( {q}^
{4}+{q}^{2}+1-{q}^{3}-q \right) }} \nn \\ W_{{5,5}}={\frac { \left( {q}^{6}+{
q}^{3}+1 \right) q}{ \left( {q}^{4}+{q}^{2}+1+{q}^{3}+q \right)
 \left( {q}^{4}+{q}^{2}+1-{q}^{3}-q \right) }} \nn \\ W_{{6,5}}=-1/2\,{\frac
{\sqrt {{q}^{6}+{q}^{5}+{q}^{4}+{q}^{3}+{q}^{2}+q+1}\sqrt {{q}^{6}-{q}
^{5}+{q}^{4}-{q}^{3}+{q}^{2}-q+1} \left( {q}^{2}+1 \right) q}{\sqrt {{
q}^{4}+{q}^{2}+1+{q}^{3}+q}\sqrt {{q}^{4}+{q}^{2}+1-{q}^{3}-q}\sqrt {{
q}^{12}+2\,{q}^{10}+4\,{q}^{8}+5\,{q}^{6}+4\,{q}^{4}+2\,{q}^{2}+1}}} \nn \\ W
_{{7,5}}=-1/2\,{\frac { \left( {q}^{6}+{q}^{5}+{q}^{4}+{q}^{3}+{q}^{2}
+q+1 \right)  \left( {q}^{6}-{q}^{5}+{q}^{4}-{q}^{3}+{q}^{2}-q+1
 \right) }{\sqrt {{q}^{4}+{q}^{2}+1+{q}^{3}+q}\sqrt {{q}^{2}+q+1}
\sqrt {{q}^{2}-q+1}\sqrt {{q}^{4}+{q}^{2}+1-{q}^{3}-q}\sqrt {{q}^{12}+
2\,{q}^{10}+4\,{q}^{8}+5\,{q}^{6}+4\,{q}^{4}+2\,{q}^{2}+1}}} \nn \\ W_{{8,5}}
=1/2\,{\frac {\sqrt {2}q\sqrt {{q}^{6}-{q}^{5}+{q}^{4}-{q}^{3}+{q}^{2}
-q+1}\sqrt {{q}^{6}+{q}^{5}+{q}^{4}+{q}^{3}+{q}^{2}+q+1}}{\sqrt {{q}^{
4}+{q}^{2}+1+{q}^{3}+q}\sqrt {{q}^{2}-q+1} \left( {q}^{2}+1 \right)
\sqrt {{q}^{2}+q+1}\sqrt {{q}^{4}-{q}^{2}+1}}} \nn \\ W_{{9,5}}=-1/2\,{\frac
{\sqrt {2}{q}^{2}\sqrt {{q}^{6}+{q}^{5}+{q}^{4}+{q}^{3}+{q}^{2}+q+1}
\sqrt {{q}^{6}-{q}^{5}+{q}^{4}-{q}^{3}+{q}^{2}-q+1}}{ \left( {q}^{4}+{
q}^{2}+1+{q}^{3}+q \right)  \left( {q}^{2}+1 \right) \sqrt {{q}^{4}+{q
}^{2}+1-{q}^{3}-q}\sqrt {{q}^{4}+1}}} \nn
\ee
\be
W_{{1,6}}=1/2\,{\frac {q \left( {q}^{2}-q+1 \right)  \left( {q}^{2}+q
+1 \right) ^{2}\sqrt {2}\sqrt {{q}^{6}+{q}^{5}+{q}^{4}+{q}^{3}+{q}^{2}
+q+1}\sqrt {{q}^{6}-{q}^{5}+{q}^{4}-{q}^{3}+{q}^{2}-q+1}}{ \left( {q}^
{4}+{q}^{2}+1+{q}^{3}+q \right)  \left( {q}^{2}+1 \right) \sqrt {{q}^{
4}+{q}^{2}+1-{q}^{3}-q}\sqrt {{q}^{12}+2\,{q}^{10}+4\,{q}^{8}+5\,{q}^{
6}+4\,{q}^{4}+2\,{q}^{2}+1}}} \nn \\ W_{{2,6}}=1/2\,{\frac {\sqrt {{q}^{4}+{q
}^{2}+1-{q}^{3}-q}\sqrt {{q}^{6}+{q}^{5}+{q}^{4}+{q}^{3}+{q}^{2}+q+1}
\sqrt {{q}^{6}-{q}^{5}+{q}^{4}-{q}^{3}+{q}^{2}-q+1}\sqrt {{q}^{4}+1}
\sqrt {{q}^{2}+q+1}\sqrt {2}}{\sqrt {{q}^{4}-{q}^{2}+1}\sqrt {{q}^{2}-
q+1} \left( {q}^{2}+1 \right) \sqrt {{q}^{12}+2\,{q}^{10}+4\,{q}^{8}+5
\,{q}^{6}+4\,{q}^{4}+2\,{q}^{2}+1}}} \nn \\ W_{{3,6}}=-1/2\,{\frac {\sqrt {{q
}^{6}+{q}^{5}+{q}^{4}+{q}^{3}+{q}^{2}+q+1}\sqrt {{q}^{6}-{q}^{5}+{q}^{
4}-{q}^{3}+{q}^{2}-q+1}{q}^{3}\sqrt {2}}{\sqrt {{q}^{4}+1}\sqrt {{q}^{
4}+{q}^{2}+1+{q}^{3}+q} \left( {q}^{2}+1 \right) \sqrt {{q}^{12}+2\,{q
}^{10}+4\,{q}^{8}+5\,{q}^{6}+4\,{q}^{4}+2\,{q}^{2}+1}}} \nn \\ W_{{4,6}}=1/2
\,{\frac { \left( {q}^{10}-2\,{q}^{9}+2\,{q}^{8}-3\,{q}^{7}+2\,{q}^{6}
-3\,{q}^{5}+2\,{q}^{4}-3\,{q}^{3}+2\,{q}^{2}-2\,q+1 \right) \sqrt {{q}
^{2}+q+1}}{\sqrt {{q}^{4}+{q}^{2}+1+{q}^{3}+q}\sqrt {{q}^{4}+{q}^{2}+1
-{q}^{3}-q}\sqrt {{q}^{2}-q+1}\sqrt {{q}^{12}+2\,{q}^{10}+4\,{q}^{8}+5
\,{q}^{6}+4\,{q}^{4}+2\,{q}^{2}+1}}} \nn \\ W_{{5,6}}=-1/2\,{\frac {\sqrt {{q
}^{6}+{q}^{5}+{q}^{4}+{q}^{3}+{q}^{2}+q+1}\sqrt {{q}^{6}-{q}^{5}+{q}^{
4}-{q}^{3}+{q}^{2}-q+1} \left( {q}^{2}+1 \right) q}{\sqrt {{q}^{4}+{q}
^{2}+1+{q}^{3}+q}\sqrt {{q}^{4}+{q}^{2}+1-{q}^{3}-q}\sqrt {{q}^{12}+2
\,{q}^{10}+4\,{q}^{8}+5\,{q}^{6}+4\,{q}^{4}+2\,{q}^{2}+1}}} \nn \\ W_{{6,6}}=
-1/2\,{\frac {{q}^{2} \left( 1-2\,{q}^{3}-4\,{q}^{7}-4\,{q}^{5}-{q}^{4
}-2\,{q}^{9}-{q}^{8}-2\,{q}^{6}+{q}^{12} \right) }{ \left( {q}^{4}+{q}
^{2}+1+{q}^{3}+q \right)  \left( {q}^{12}+2\,{q}^{10}+4\,{q}^{8}+5\,{q
}^{6}+4\,{q}^{4}+2\,{q}^{2}+1 \right) }} \nn \\ W_{{7,6}}=1/2\,{\frac {
 \left( {q}^{6}-{q}^{5}+{q}^{4}+{q}^{3}+{q}^{2}-q+1 \right) \sqrt {{q}
^{2}+q+1}\sqrt {{q}^{6}+{q}^{5}+{q}^{4}+{q}^{3}+{q}^{2}+q+1}\sqrt {{q}
^{6}-{q}^{5}+{q}^{4}-{q}^{3}+{q}^{2}-q+1}}{ \left( {q}^{12}+2\,{q}^{10
}+4\,{q}^{8}+5\,{q}^{6}+4\,{q}^{4}+2\,{q}^{2}+1 \right) \sqrt {{q}^{2}
-q+1}}} \nn \\ W_{{8,6}}=-1/2\,{\frac {\sqrt {{q}^{4}+{q}^{2}+1-{q}^{3}-q}
 \left( {q}^{6}+{q}^{3}+1 \right) q\sqrt {2}\sqrt {{q}^{2}+q+1}}{
\sqrt {{q}^{4}-{q}^{2}+1}\sqrt {{q}^{2}-q+1} \left( {q}^{2}+1 \right)
\sqrt {{q}^{12}+2\,{q}^{10}+4\,{q}^{8}+5\,{q}^{6}+4\,{q}^{4}+2\,{q}^{2
}+1}}} \nn \\ W_{{9,6}}=1/2\,{\frac {\sqrt {2}{q}^{2} \left( {q}^{8}+{q}^{7}+
2\,{q}^{6}+{q}^{5}+3\,{q}^{4}+{q}^{3}+2\,{q}^{2}+q+1 \right) }{\sqrt {
{q}^{4}+1}\sqrt {{q}^{4}+{q}^{2}+1+{q}^{3}+q} \left( {q}^{2}+1
 \right) \sqrt {{q}^{12}+2\,{q}^{10}+4\,{q}^{8}+5\,{q}^{6}+4\,{q}^{4}+
2\,{q}^{2}+1}}} \nn
\ee
\be
W_{{1,7}}=-1/2\,{\frac {\sqrt {2}{q}^{3} \left( {q}^{6}+{q}^{5}+{q}^{
4}+{q}^{3}+{q}^{2}+q+1 \right) }{\sqrt {{q}^{12}+2\,{q}^{10}+4\,{q}^{8
}+5\,{q}^{6}+4\,{q}^{4}+2\,{q}^{2}+1} \left( {q}^{2}+1 \right) \sqrt {
{q}^{2}-q+1}\sqrt {{q}^{2}+q+1}\sqrt {{q}^{4}+{q}^{2}+1-{q}^{3}-q}}} \nn \\ W
_{{2,7}}=-1/2\,{\frac { \left( {q}^{8}-{q}^{7}+{q}^{6}-2\,{q}^{5}+{q}^
{4}-2\,{q}^{3}+{q}^{2}-q+1 \right) {q}^{3}\sqrt {{q}^{4}+1}\sqrt {2}}{
\sqrt {{q}^{4}-{q}^{2}+1} \left( {q}^{2}+q+1 \right)  \left( {q}^{2}-q
+1 \right)  \left( {q}^{2}+1 \right) \sqrt {{q}^{4}+{q}^{2}+1-{q}^{3}-
q}\sqrt {{q}^{12}+2\,{q}^{10}+4\,{q}^{8}+5\,{q}^{6}+4\,{q}^{4}+2\,{q}^
{2}+1}}} \nn \\ W_{{3,7}}=1/2\,{\frac { \left( {q}^{8}-{q}^{7}+2\,{q}^{6}-{q}
^{5}+3\,{q}^{4}-{q}^{3}+2\,{q}^{2}-q+1 \right) q\sqrt {2}\sqrt {{q}^{4
}+{q}^{2}+1+{q}^{3}+q}}{\sqrt {{q}^{2}+q+1}\sqrt {{q}^{2}-q+1}\sqrt {{
q}^{4}+1} \left( {q}^{2}+1 \right) \sqrt {{q}^{12}+2\,{q}^{10}+4\,{q}^
{8}+5\,{q}^{6}+4\,{q}^{4}+2\,{q}^{2}+1}}} \nn \\ W_{{4,7}}=-1/2\,{\frac {
 \left( {q}^{2}+1 \right)  \left( {q}^{6}+{q}^{5}+{q}^{4}+{q}^{3}+{q}^
{2}+q+1 \right) ^{3/2}\sqrt {{q}^{6}-{q}^{5}+{q}^{4}-{q}^{3}+{q}^{2}-q
+1}}{ \left( {q}^{2}-q+1 \right)  \left( {q}^{2}+q+1 \right) \sqrt {{q
}^{4}+{q}^{2}+1+{q}^{3}+q}\sqrt {{q}^{4}+{q}^{2}+1-{q}^{3}-q}\sqrt {{q
}^{12}+2\,{q}^{10}+4\,{q}^{8}+5\,{q}^{6}+4\,{q}^{4}+2\,{q}^{2}+1}}} \nn \\ W_
{{5,7}}=-1/2\,{\frac { \left( {q}^{6}+{q}^{5}+{q}^{4}+{q}^{3}+{q}^{2}+
q+1 \right)  \left( {q}^{6}-{q}^{5}+{q}^{4}-{q}^{3}+{q}^{2}-q+1
 \right) }{\sqrt {{q}^{4}+{q}^{2}+1+{q}^{3}+q}\sqrt {{q}^{2}+q+1}
\sqrt {{q}^{2}-q+1}\sqrt {{q}^{4}+{q}^{2}+1-{q}^{3}-q}\sqrt {{q}^{12}+
2\,{q}^{10}+4\,{q}^{8}+5\,{q}^{6}+4\,{q}^{4}+2\,{q}^{2}+1}}} \nn \\ W_{{6,7}}
=1/2\,{\frac { \left( {q}^{6}-{q}^{5}+{q}^{4}+{q}^{3}+{q}^{2}-q+1
 \right) \sqrt {{q}^{2}+q+1}\sqrt {{q}^{6}+{q}^{5}+{q}^{4}+{q}^{3}+{q}
^{2}+q+1}\sqrt {{q}^{6}-{q}^{5}+{q}^{4}-{q}^{3}+{q}^{2}-q+1}}{ \left(
{q}^{12}+2\,{q}^{10}+4\,{q}^{8}+5\,{q}^{6}+4\,{q}^{4}+2\,{q}^{2}+1
 \right) \sqrt {{q}^{2}-q+1}}} \nn \\ W_{{7,7}}=1/2\,{\frac {[11]-2[10]+4[9]-6[8]+10[7]-12[6]+14[5]-13[4]+15[3]-11[2]+6
}{ \left( {q}^{4}+{q}^{2}+1-{q}^{3}-q \right)  \left( {q}^{
2}+q+1 \right)  \left( {q}^{2}-q+1 \right)  \left( {q}^{12}+2\,{q}^{10
}+4\,{q}^{8}+5\,{q}^{6}+4\,{q}^{4}+2\,{q}^{2}+1 \right) }} \nn \\ W_{{8,7}}=1
/2\,{\frac {\sqrt {2}{q}^{2}\sqrt {{q}^{6}+{q}^{5}+{q}^{4}+{q}^{3}+{q}
^{2}+q+1}\sqrt {{q}^{6}-{q}^{5}+{q}^{4}-{q}^{3}+{q}^{2}-q+1} \left( {q
}^{6}+2\,{q}^{4}-{q}^{3}+2\,{q}^{2}+1 \right) }{\sqrt {{q}^{4}-{q}^{2}
+1} \left( {q}^{2}+q+1 \right)  \left( {q}^{2}-q+1 \right)  \left( {q}
^{2}+1 \right) \sqrt {{q}^{4}+{q}^{2}+1-{q}^{3}-q}\sqrt {{q}^{12}+2\,{
q}^{10}+4\,{q}^{8}+5\,{q}^{6}+4\,{q}^{4}+2\,{q}^{2}+1}}} \nn \\ W_{{9,7}}=-1/
2\,{\frac {{q}^{4}\sqrt {2}\sqrt {{q}^{6}+{q}^{5}+{q}^{4}+{q}^{3}+{q}^
{2}+q+1}\sqrt {{q}^{6}-{q}^{5}+{q}^{4}-{q}^{3}+{q}^{2}-q+1}\sqrt {{q}^
{2}+q+1}\sqrt {{q}^{2}-q+1}}{ \left( {q}^{4}+{q}^{2}+1-{q}^{3}-q
 \right) \sqrt {{q}^{4}+{q}^{2}+1+{q}^{3}+q} \left( {q}^{2}+1 \right)
\sqrt {{q}^{4}+1}\sqrt {{q}^{12}+2\,{q}^{10}+4\,{q}^{8}+5\,{q}^{6}+4\,
{q}^{4}+2\,{q}^{2}+1}}} \nn
\ee
}

For transposed diagrams
\be
{\cal R}_{\tilde Q}(q) = {\cal R}_Q (-q^{-1}), \ \ \ \ \ \ \
{\cal U}_{\tilde Q}(q) = {\cal U}_Q(-q^{-1})
\ee
Minus sign is important: for example,
\be
{\cal R}_{[2,2,2,1,1,1]} = -q^{\varkappa_{[2,2,1,1]}} = -q^{5} = {\cal R}_{[6,3]}(-q^{-1})
\ee
and
\be
{\cal R}_{[4,1,1,1,1,1]} = q^{\varkappa_{[3,1,1]}} = q^{3}= {\cal R}_{[6,1,1]}(-q^{-1})
\ee

The list of transpositions:
{\footnotesize
\be
\begin{array}{c|c|c}
\text{matrix size} & \tilde Q & Q \\
&&\\ \hline && \\
1 & [4,1,1,1,1,1] & [6,1,1,1]\\
 & [2,2,2,1,1,1] & [6,3] \\
 &&\\ \hline && \\
2 & [3,2,1,1,1,1]& [6,2,1] \\
& [2,2,2,2,1] & [5,4] \\
& [5,1,1,1,1] & \text{symmetric} \\
& [3,3,3] & \text{symmetric} \\
 &&\\ \hline && \\
4 & [3,3,1,1,1] & [5,2,2] \\
& [3,2,2,2] & [4,4,1] \\
 &&\\ \hline && \\
6 & [4,2,1,1,1] & [5,2,1,1] \\
& [3,2,2,1,1] & [5,3,1] \\
 &&\\ \hline && \\
8 & [3,3,2,1] & [4,3,2] \\
&&\\ \hline && \\
9 & [4,2,2,1] & [4,3,1,1] \\ &&
 \end{array}
 \ee
}

\section*{Appendix B. Matrices $T,\bar T$ and $S,\bar S$}

In this Appendix we list the matrices $T,\bar T$ and $S,\bar S$ in representation $[2,1]$ necessary for evaluating the $[2,1]$-colored HOMFLY polynomials through the formulas of s.4, see \cite{mmmrs}. Due to participating the conjugated representation $\overline{[2,1]}$, all the quantities depend on the group $SU_q(N)$.

In this case there are seven different items $X$ in the decomposition
\be
\l[2,1]\otimes \underbrace{\overline{[2,1]}}_{[2^{N-2},1]} = [4,3,2^{N-4},1]\oplus[4,2^{N-2}]
\oplus[4,2^{N-3},1,1]\oplus[3,3,2^{N-3}]\oplus[3,3,2^{N-4},1,1]\oplus \underline{2\cdot [3,2^{N-2},1]}\oplus \underbrace{[2^N]}_{singlet}
\label{21bar21}
\ee
Only one (underlined) item $[3,2^{N-2},1]=[2,1^{N-2}]=Adj$ enters (\ref{21bar21})
with non-trivial multiplicity $m_{Adj}=2$,
however this makes the Racah matrices $10\times 10$, since $\sum_X m_X^2 = 6\cdot 1+1\cdot 2^2 = 10$.

The corresponding dimensions and eigenvalues are
\be
\begin{array}{c|c|c}
X & d_X & t_X \\
&& \\
\hline
&& \\
\l[2^N]   &1 & 1\\
&& \\
\l[4,2^{N-3},1,1]&\frac{[N-2][N-1][N+1][N+2]}{[2]^2}& -A^2\\
&& \\
\l[3,3,2^{N-3}]&\frac{[N-2][N-1][N+1][N+2]}{[2]^2}& -A^2 \\
&&\\
\l[3,3,2^{N-4},1,1]&\frac{[N-3][N]^2[N+1]}{[2]^2}& - q^{-2}A^2\\
&& \\
\l[4,3,2^{N-4},1]&\frac{[N-3][N-1]^2[N+1]^2[N+3]}{[3]^2}& A^3 \\
&& \\
\l[4,2^{N-2}] &\frac{[N-1][N]^2[N+3]}{[2]^2}& q^2A^2\\
&& \\
\hline
&& \\
\underline{\l[3,2^{N-2},1]}&\underline{[N-1][N+1]} & \underline{\pm A} \\
&& \\
\end{array}
\label{dims21a}
\ee
Similarly in order to construct matrix $S$, one also needs the expansion (\ref{212}) and
\be
\begin{array}{c|c|c}
X & d_X & t_X \\
&& \\
\hline
&& \\
\l[2,2,2]&\frac{[N-2][N-1]^2[N]^2[N+1]}{[2]^2[3]^2[4]}& q^{-3}A^{-3}\\
&& \\
\l[4,1,1] &\frac{[N-2][N-1][N][N+1][N+2][N+3]}{[2]^2[3][6]}& -q^3A^{-3}\\
&& \\
\l[3,3]&\frac{[N-1][N]^2[N+1]^2[N+2]}{[2]^2[3]^2[4]}& -q^3A^{-3}\\
&& \\
\l[2,2,1,1]   &\frac{[N-3][N-2][N-1][N]^2[N+1]}{[2]^2[4][5]} & -q^{-5}A^{-3}\\
&&\\
 \l[3,1,1,1]&\frac{[N-3][N-2][N-1][N][N+1][N+2]}{[2]^2[3][6]} & q^{-3}A^{-3} \\
&& \\
\l[4,2]&\frac{[N-1][N]^2[N+1][N+2][N+3]}{[2]^2[4][5]}&q^5A^{-3}   \\
&& \\
\hline
&& \\
\underline{\l[3,2,1]}&\underline{\frac{[N-2][N-1][N]^2[N+1][N+2]}{[3]^2[5]}}&\underline{\pm A^{-3}} \\
&&\\
\end{array}
\label{dims21p}
\ee
Here we again encounter one item with multiplicity two. The diagonal matrices $\bar T$ and $T$ are read off from the last columns of
(\ref{dims21a}) and (\ref{dims21p}):
\be
\overline T = \left(\begin{array}{cccccc|cccc}
1&&&&&&&&&\\
&-A^2&&&&&&&&\\
&&-A^2&&&&&&&\\
&&&-q^{-2}A^2&&&&&&\\
&&&&A^3&&&&&\\
&&&&&q^2A^2&&&&\\
\hline
&&&&&&A&&&\\
&&&&&&&-A&&\\
&&&&&&&&A&\\
&&&&&&&&& -A
\end{array}\right)
\label{Ta21}
\ee
and
\be
 T = \frac{1}{A^3}\left(\begin{array}{cccccc|cccc}
q^{-3}&&&&&&&&&\\
&-q^3&&&&&&&&\\
&&-q^3&&&&&&&\\
&&&-q^{-5}&&&&&&\\
&&&&q^{-3}&&&&&\\
&&&&&q^5&&&&\\
\hline
&&&&&&1&&&\\
&&&&&&&-1&&\\
&&&&&&&&-1&\\
&&&&&&&&& 1
\end{array}\right)
\label{Tp21}
\ee
while the matrices $\bar S$ and $S$ are respectively

\begin{landscape}
\tiny{
\be
\left(\begin{array}{cccccc|cccc}
\frac{[3]}{D_{-1}D_0D_1}&{[3]\over [2]D_0}\sqrt{D_2D_{-2}\over D_1D_{-1}}&{[3]\over [2]D_0}\sqrt{D_2D_{-2}\over D_1D_{-1}}
&{[3]\over [2]D_{-1}}\sqrt{D_{-3}\over D_1}&{\sqrt{D_3D_{-3}}\over D_0}&{[3]\over [2]D_1}\sqrt{D_3\over D_{-1}}&
{[3]\over D_0}{1\over\sqrt{D_1D_{-1}}}&{[3]\over D_0}{1\over\sqrt{D_1D_{-1}}}&0&0\\
&&&&&&&&&\\
{[3]\over [2]D_0}\sqrt{D_2D_{-2}\over D_1D_{-1}}&{[3]\over [2]^2D_0}&{[3]\over [2]^2D_0}&
-{[3]\over [2]^2D_0}\sqrt{D_{-3}D_2\over D_{-2}D_{-1}}&0&-{[3]\over [2]^2D_0}\sqrt{D_3D_{-2}\over D_2D_1}&\bar S_{[2,7]}&
-{\sqrt{D_2D_{-2}}\over [2]D_0}&{D_{-1}\over [2]D_0}&{D_1\over [2]D_0}\\
&&&&&&&&&\\
{[3]\over [2]D_0}\sqrt{D_2D_{-2}\over D_1D_{-1}}&{[3]\over [2]^2D_0}&{[3]\over [2]^2D_0}&-{[3]\over [2]^2D_0}
\sqrt{D_{-3}D_2\over D_{-1}D_{-2}}&0&-{[3]\over [2]^2D_0}
\sqrt{D_3D_{-2}\over D_1D_2}&\bar S_{[2,7]}&-{\sqrt{D_2D_{-2}}\over [2]D_0}&-{D_1\over [2]D_0}&-{D_{-1}\over [2]D_0}\\
&&&&&&&&&\\
{[3]\over [2]D_{-1}}\sqrt{D_3\over D_1}&-{[3]\over [2]^2D_0}\sqrt{D_{-3}D_2\over D_{-2}D_{-1}}&-{[3]\over [2]^2D_0}
\sqrt{D_{-3}D_2\over D_{-1}D_{-2}}&\bar S_{[4,4]}&{[3]\sqrt{D_1D_3}\over D_0D_2D_{-2}}&-{[3]^2D_0\over D_2D_{-2}}
\sqrt{D_3D_{-3}\over D_1D_{-1}}&\bar S_{[4,7]}&{D_2\over [2]D_0}\sqrt{D_{-3}\over D_{-1}}&
-{\sqrt{D_{-1}D_{-3}D_2}\over [2]D_0\sqrt{D_{-2}}}&{\sqrt{D_{-1}D_{-3}D_2}\over [2]D_0\sqrt{D_{-2}}}\\
&&&&&&&&&\\
{\sqrt{D_3D_{-3}}\over D_0}&0&0&{[3]\sqrt{D_1D_3}\over D_0D_2D_{-2}}&-{[3]\over D_{-2}D_0D_2}&
{[3]\sqrt{D_{-1}D_{-3}}\over D_{-2}D_0D_2}&-{[3]\sqrt{D_{-1}D_1D_{-3}D_3}\over D_{-2}D_0D_2}&0&0&0\\
&&&&&&&&&\\
{[3]\over [2]D_1}\sqrt{D_3\over D_{-1}}&-{[3]\over [2]^2D_0}\sqrt{D_3D_{-2}\over D_2D_1}&-{[3]\over [2]^2D_0}
\sqrt{D_3D_{-2}\over D_1D_2}&-{[3]^2D_0\over D_2D_{-2}}
\sqrt{D_3D_{-3}\over D_1D_{-1}}&{[3]\sqrt{D_{-1}D_{-3}}\over D_{-2}D_0D_2}&\bar S_{[6,6]}&\bar S_{[6,7]}&
{D_{-2}\over [2]D_0}\sqrt{D_3\over D_1}&{\sqrt{D_1D_3D_{-2}}\over [2]D_0\sqrt{D_{2}}}&
-{\sqrt{D_1D_3D_{-2}}\over [2]D_0\sqrt{D_{2}}}
\\&&&&&&&&&\\
\hline
&&&&&&&&&\\
{[3]\over D_0}{1\over\sqrt{D_1D_{-1}}}&\bar S_{[2,7]}&\bar S_{[2,7]}&\bar S_{[4,7]}&
-{[3]\sqrt{D_{-1}D_1D_{-3}D_3}\over D_{-2}D_0D_2}&\bar S_{[6,7]}&\bar S_{[7,7]}&{[4]\over [2]D_0}&
-{\{A^2\}\over \{q\}}{1\over D_0^2\sqrt{D_2D_{-2}}}&{\{A^2\}\over \{q\}}{1\over D_0^2\sqrt{D_2D_{-2}}}\\
&&&&&&&&&\\
{[3]\over D_0}{1\over\sqrt{D_1D_{-1}}}&-{\sqrt{D_2D_{-2}}\over [2]D_0}&-{\sqrt{D_2D_{-2}}\over [2]D_0}&
{D_2\over [2]D_0}\sqrt{D_{-3}\over D_{-1}}&0&{D_{-2}\over [2]D_0}\sqrt{D_3\over D_1}&{[4]\over [2]D_0}&-{1\over D_0}&
0&0\\
&&&&&&&&&\\
0&-{D_1\over [2]D_0}&{D_{-1}\over [2]D_0}&-{\sqrt{D_{-1}D_{-3}D_2}\over [2]D_0\sqrt{D_{-2}}}&0&
{\sqrt{D_1D_3D_{-2}}\over [2]D_0\sqrt{D_{2}}}&-{\{A^2\}\over \{q\}}{1\over D_0^2\sqrt{D_2D_{-2}}}&0&-{1\over D_0}&
{1\over D_0}\\
&&&&&&&&&\\
0&-{D_{-1}\over [2]D_0}&{D_1\over [2]D_0}&{\sqrt{D_{-1}D_{-3}D_2}\over [2]D_0\sqrt{D_{-2}}}&0&
-{\sqrt{D_1D_3D_{-2}}\over [2]D_0\sqrt{D_{2}}}&{\{A^2\}\over \{q\}}{1\over D_0^2\sqrt{D_2D_{-2}}}&0&{1\over D_0}&-{1\over D_0}
\end{array}\right)
\nn
\ee
}

\bigskip

{\footnotesize
\be
\bar S_{[2,7]}=
-{[2][3]^2D_{-2}D_0D_2-[2]D_2^2D_0D_{-2}^2-[3]^3D_0(D_2D_{-3}+D_3D_{-2})+[3]D_2D_{-2}(D_3+D_{-3})\over
[2]^2D_0^2D_1D_{-1}\sqrt{D_2D_{-2}}}
\nn
\ee
\be
\bar S_{[4,4]}={[2]^2(D_1+D_{-1})-D_{-3}D_0D_{4}\over [2]^2D_0D_{-1}D_{-2}D_2}
\nn
\ee
\be
\bar S_{[4,7]}=\sqrt{D_{-3}\over D_{-1}^3}
{[2]^2[3]^2+2[3]D_2D_{-2}(D_{-1}^2+D_1^2)-[3]^2D_0D_1(D_2D_{-3}+D_3D_{-2})-[2]D_{-1}D_{-2}^2D_2^3\over [2]^2D_0D_2D_{-2}D_1^2}
\label{Sa21}
\ee
\be
\bar S_{[6,7]}=\sqrt{D_{3}\over D_{1}^3}
{[2]^2[3]^2+2[3]D_2D_{-2}(D_{-1}^2+D_1^2)-[3]^2D_0D_{-1}(D_2D_{-3}+D_3D_{-2})-[2]D_{1}D_{2}^2D_{-2}^3\over [2]^2D_0D_2D_{-2}D_{-1}^2}
\nn
\ee
\be
\bar S_{[6,6]}={[2]^2(D_1+D_{-1})-D_3D_0D_{-4}\over [2]^2D_0D_1D_2D_{-2}}
\nn
\ee
\be
\bar S_{[7,7]}=
{[2]D_{-2}D_2(D_2D_{-1}^2+D_{-2}D_1^2)-2[3](D_{-1}^3+D_1^3))\over [2]D_0^2D_{-1}^2D_1^2}+
{[2][3]^2\{A^2\}^2\over \{q\}^2D_{-2}D_2D_0^3D_1^2D_{-1}^2(D_1D_{-2}+D_{-1}D_2)}+
{[3]^2(D_2D_{-3}+D_3D_{-2})\over D_{-2}D_0D_2(D_1D_{-2}+D_{-1}D_2)}
\nn
\ee}
\end{landscape}

\bigskip

\begin{landscape}
\small{
\be
\left(\begin{array}{cccccc|cccc}
\sqrt{\frac{ D_{-2}}{[2]^2[4]\,D_1 }} &\sqrt{\frac{[3]\,D_3D_2D_{-2}}{[2]^2[6]\,D_1D_0D_{-1} }}
&\sqrt{\frac{ D_2}{[2]^2[4]\,D_{-1} }}
&\sqrt{\frac{ [3]^2\,D_{-2}D_{-3}}{[2]^2[4][5]\,D_1D_{-1} }}
&\sqrt{\frac{ [3]\,D_2D_{-2}D_{-3}}{[2]^2[6]\,D_1D_0D_{-1} }}
&\sqrt{\frac{ [3]^2D_3D_2}{[2]^2[4][5]\,D_1D_{-1} }}
&\sqrt{\frac{D_2D_{-2}}{[5]\,D_1D_{-1} }}&\sqrt{\frac{D_2D_{-2}}{[5]\,D_1D_{-1}}}&0&0\\
&&&&&&&&&\\
{[3]\sqrt{D_{-1}D_2}\over [2]^2\sqrt{[4]}D_0}&{\sqrt{[3]D_3}\over [2]^2\sqrt{[6]D_0}}&{[3]\sqrt{D_1D_{-2}}\over [2]^2D_0\sqrt{[4]}}
&-{[3]\sqrt{D_2D_{-3}}\over [2]^2D_0\sqrt{[4][5]}}&{\sqrt{[3]D_{-3}}\over [2]^2\sqrt{D_0[6]}}&-{[3]\sqrt{D_3D_{-2}}\over
[2]^2D_0\sqrt{[4][5]}}&-{[4]D_0+[3](D_2-D_{-2})\over 2[2]^2\sqrt{[5]}D_0}
&-{[4]D_0-[3](D_2-D_{-2})\over 2[2]^2\sqrt{[5]}D_0}&\frac{1}{2}&\frac{1}{2}\\
&&&&&&&&&\\
{[3]\sqrt{D_{-1}D_2}\over [2]^2\sqrt{[4]}D_0}&{\sqrt{[3]D_3}\over [2]^2\sqrt{[6]D_0}}&{[3]\sqrt{D_1D_{-2}}\over [2]^2D_0\sqrt{[4]}}
&-{[3]\sqrt{D_2D_{-3}}\over [2]^2D_0\sqrt{[4][5]}}&{\sqrt{[3]D_{-3}}\over [2]^2\sqrt{D_0[6]}}&-{[3]\sqrt{D_3D_{-2}}\over
[2]^2D_0\sqrt{[4][5]}}&-{[4]D_0+[3](D_2-D_{-2})\over 2[2]^2\sqrt{[5]}D_0}
&-{[4]D_0-[3](D_2-D_{-2})\over 2[2]^2\sqrt{[5]}D_0}&-\frac{1}{2}&-\frac{1}{2}\\
&&&&&&&&&\\
{\sqrt{D_{-3}}\over [2]^2\sqrt{[4]}\sqrt{D_{-2}}}&{\sqrt{D_0D_3D_{-3}[3]^3}\over [2]^2\sqrt{[6]D_{-1}D_{-2}D_2}}&-{[3]\sqrt{D_1D_{-3}}
\over [2]^2\sqrt{[4]D_2D_{-1}}}
&-{[4]D_{-1}+[2][3]D_1\over [2]^2\sqrt{[4][5]D_{-1}D_{-2}}}&S_{[4,5]}&-{[3]^2D_3D_{-3}\over [2]^2\sqrt{[4][5]D_2D_{-1}}}
&{\sqrt{D_{-3}}(D_2+D_{-1})\over [2]\sqrt{[5]D_{-1}D_{-2}D_2}}&{\sqrt{D_{-3}}(D_2-D_{-1})\over [2]\sqrt{[5]D_{-1}D_{-2}D_2}}&0&0\\
&&&&&&&&&\\
{\sqrt{D_1D_3D_{-3}}\over [2]D_0\sqrt{[4]D_{-2}}}&-{\sqrt{[3]D_{-3}D_{-1}D_1}\over
 [2]\sqrt{[6]D_0D_2D_{-2}}}&-{\sqrt{D_{-3}D_{-1}D_3}\over [2]D_0\sqrt{[4]D_2}}
&-{\sqrt{[3]D_3D_1D_{-1}}\over [2]D_0\sqrt{[4][5]D_{-2}}}&{\sqrt{[3]D_1D_{-1}D_3}\over [2]\sqrt{[6]D_0D_2D_{-2}}}&
-{\sqrt{[3]D_1D_{-1}D_{-3}}\over [2]D_0\sqrt{[4][5]D_2}}
&{\sqrt{D_1D_{-1}D_3D_{-3}}\over D_0\sqrt{[5]D_2D_{-2}}}&{\sqrt{D_1D_{-1}D_3D_{-3}}\over D_0\sqrt{[5]D_2D_{-2}}}&0&0\\
&&&&&&&&&\\
{[3]\sqrt{D_{-1}D_3}\over [2]^2\sqrt{[4]D_1D_{-2}}}&S_{[6,2]}&-{\sqrt{D_3}\over [2]^2\sqrt{[4]D_2}}
&{[3]^2\sqrt{D_3D_{-3}}\over [2]^2\sqrt{[4][5]D_1D_{-2}}}&-{\sqrt{D_0D_3D_{-3}[3]^3}\over [2]^2\sqrt{[6]D_1D_2D_{-2}}}
&{[2][3]D_{-1}+[4]D_1\over [2]^3\sqrt{[4][5]D_1D_2}}
&-{\sqrt{D_3}(D_{-2}-D_1)\over [2]\sqrt{[5]D_1D_2D_{-2}}}&-{\sqrt{D_3}(D_{-2}+D_1)\over [2]\sqrt{[5]D_1D_2D_{-2}}}&0&0\\
&&&&&&&&&\\
\hline
&&&&&&&&&\\
-{\sqrt{D_{-1}^3}\over [2]D_0\sqrt{[4]D_{-2}}}&S_{[7,2]}&{\sqrt{D_{1}^3}\over [2]D_0\sqrt{[4]D_{2}}}
&-{\sqrt{D_{-3}}([2][3]D_{1}+[4]D_{-1})\over [2]^2D_0\sqrt{[4][5]D_{-2}}}&S_{[7,5]}
&{\sqrt{D_{3}}([2][3]D_{-1}+[4]D_{1}\over [2]^2D_0)\sqrt{[4][5]D_{2}}}&-{{\{A^2\}\over\{q\}}+1\over D_0\sqrt{[5]D_2D_{-2}}}
&-{{\{A^2\}\over\{q\}}-1\over D_0\sqrt{[5]D_2D_{-2}}}&0&0\\
&&&&&&&&&\\
-{\sqrt{D_{-1}D_{-2}}\over D_0\sqrt{[4]}}&0&{\sqrt{D_{1}D_{2}}\over D_0\sqrt{[4]}}
&{\sqrt{D_{-3}D_{-2}}\over D_0\sqrt{[4][5]}}&0&-{\sqrt{D_{3}D_{2}}\over D_0\sqrt{[4][5]}}
&{\sqrt{D_{2}D_{-2}}\over D_0\sqrt{[5]}}&-{\sqrt{D_{2}D_{-2}}\over D_0\sqrt{[5]}}&0&0\\
&&&&&&&&&\\
{\sqrt{D_{-1}D_2}\over [2]D_0\sqrt{[4]}}&-{\sqrt{[3]D_3}\over [2]\sqrt{[6]D_0}}&{\sqrt{D_1D_{-2}}\over [2]D_0\sqrt{[4]}}
&-{\sqrt{D_{-3}D_2}\over [2]D_0\sqrt{[4][5]}}&-{\sqrt{[3]D_{-3}}\over [2]\sqrt{[6]D_0}}&-{\sqrt{D_3D_{-2}}\over [2]D_0\sqrt{[4][5]}}
&{[6]D_0+[3](D_{-2}-D_2)\over 2[2][3]D_0\sqrt{[5]}}&{[6]D_0+[3](D_{2}-D_{-2})\over 2[2][3]D_0\sqrt{[5]}}&{1\over 2}&-{1\over 2}\\
&&&&&&&&&\\
{\sqrt{D_{-1}D_2}\over [2]D_0\sqrt{[4]}}&-{\sqrt{[3]D_3}\over [2]\sqrt{[6]D_0}}&{\sqrt{D_1D_{-2}}\over [2]D_0\sqrt{[4]}}
&-{\sqrt{D_{-3}D_2}\over [2]D_0\sqrt{[4][5]}}&-{\sqrt{[3]D_{-3}}\over [2]\sqrt{[6]D_0}}&-{\sqrt{D_3D_{-2}}\over [2]D_0\sqrt{[4][5]}}
&{[6]D_0+[3](D_{-2}-D_2)\over 2[2][3]D_0\sqrt{[5]}}&{[6]D_0+[3](D_{2}-D_{-2})\over 2[2][3]D_0\sqrt{[5]}}&-{1\over 2}&{1\over 2}
\end{array}\right)\nn
\ee

\bigskip

\be
S_{[4,5]}={[2][3]D_{-1}D_{-2}-[6]D_{-1}D_2-[3][4]D_1D_2\over [2]^2[4]\sqrt{[3][6]D_0D_{-1}D_{-2}D_2}}
\nn
\ee
\be\label{Sp21}
S_{[6,2]}=-{[2][3]D_{1}D_{2}-[6]D_{1}D_{-2}-[3][4]D_{-1}D_{-2}\over [2]^2[4]\sqrt{[3][6]D_0D_{1}D_{-2}D_2}}
\ee
\be
S_{[7,2]}=-\sqrt{D_3}{[2][3]D_{1}D_{2}-[6]D_{1}D_{-2}-[3][4]D_{-1}D_{-2}\over [2][4]\sqrt{[3][6]D_0^3D_{-2}D_2}}
\nn
\ee
\be
S_{[7,5]}=\sqrt{D_{-3}}{[2][3]D_{-1}D_{-2}-[6]D_{-1}D_2-[3][4]D_1D_2\over [2][4]\sqrt{[3][6]D_0^3D_{-2}D_2}}
\nn
\ee
}
\end{landscape}

\end{document}